\documentstyle[epsfig,preprint,aps,eqsecnum,fixes]{revtex}

\makeatletter           
\@floatstrue
\def\figure{\let\@capwidth\columnwidth\@float{figure}}
\let\endfigure\end@float
\@namedef{figure*}{\let\@capwidth\textwidth\@dblfloat{figure}}
\@namedef{endfigure*}{\end@dblfloat}
\def\table{\let\@capwidth\columnwidth\@float{table}}
\let\endtable\end@float
\@namedef{table*}{\let\@capwidth\textwidth\@dblfloat{table}}
\@namedef{endtable*}{\end@dblfloat}

\makeatother

\tightenlines
\begin {document}

\def\Tc{T_{\rm c}}
\def\kB{k_{\rm B}}

\def\grad{\mbox{\boldmath$\nabla$}}
\def\p{{\bf p}}
\def\q{{\bf q}}
\def\k{{\bf k}}
\def\l{{\bf l}}
\def\s{{\bf s}}

\def\half{{\textstyle{1\over2}}}

\def\Li{{\rm Li}}

\def\IR{{\rm IR}}

\def\c{{\rm c}}
\def\aho{a_{\rm ho}}
\def\omegaho{\omega_{\rm ho}}
\def\omegamax{\omega_{\rm max}}
\def\threehalf{{\textstyle{3\over2}}}
\def\fivehalf{{\textstyle{5\over2}}}
\def\fourth{{\textstyle{1\over4}}}
\def\x{{\bf x}}
\def\muc{\bar\mu_{\rm c}}
\def\eff{{\rm eff}}
\def\Rnp{R_{\rm np}}
\def\Rcloud{R_{\rm cloud}}
\def\SI{S_{\rm I}}
\def\eps{\epsilon}
\def\msb{{\overline{\rm MS}}}
\def\msbar{{\overline{\scriptscriptstyle{\rm MS}}}}
\def\rms{r_\msbar^{}}
\def\MSbar{$\msb$}
\def\ren{M}
\def\Mir{{\cal N}}
\def\mir{{\cal M}}
\def\bare{{\rm bare}}
\def\gammaE{{\gamma_{\scriptscriptstyle{E}}}}
\def\sumint{\hbox{$\sum$}\!\!\!\!\!\!\int}
\def\PP{{\rm P.P.}}
\def\Re{{\rm Re}}
\def\tomega{\tilde\omega}
\def\Q{{\cal Q}}
\def\reff{r_{\rm s}}



\title
{$\Tc$ for trapped dilute Bose gases: a second-order result}

\author {Peter Arnold and Boris Tom\'{a}\v{s}ik}
\address
    {%
    Department of Physics,
    University of Virginia \\
    P.O. Box 400714,
    Charlottesville, Virginia  22904-4714
    }%
\date{\today}

\maketitle
\vskip -20pt

\begin {abstract}%
{%
   For some time, the theoretical result for the transition temperature
   of a dilute three-dimensional
   Bose gas in an arbitrarily wide harmonic trap has been
   known to first order
   in the interaction strength.  We extend that result to second order.
   The first-order result for a gas trapped in a harmonic potential
   can be computed
   in mean field theory (in contrast to the first order result for a
   uniform gas, which cannot).  We show that, at second order,
   perturbation theory suffices for relating the transition temperature
   to the chemical potential at the transition, but the chemical
   potential is non-perturbative at the desired order.
   The necessary information about the chemical potential can be extracted,
   however, from
   recent lattice simulations of uniform Bose gases.
}%
\end {abstract}

\thispagestyle{empty}


\section {Introduction}

Consider a dilute three-dimensional gas of bosons, all identical,
in an external harmonic trapping potential
\begin {equation}
   V(\x) = {1\over 2} m (\omega_x^2 x^2 + \omega_y^2 y^2 + \omega_z^2 z^2) ,
\label {eq:V}
\end {equation}
where $m$ is the mass of each boson.
For this system to have a sharp, well-defined phase transition, we need to
formally take the infinite volume limit of
$\omega_x, \omega_y, \omega_z \to 0$ while keeping the central density of
Bose particles finite and non-zero at the transition.
As we'll briefly review below, the central density $\bar n$ at the transition
scales as
$\bar n \sim N^{1/2} / \aho^3$,
where
$N$ is the total number of Bose particles in the trap, and where
\begin {equation}
   \aho^3 \equiv \left(\hbar\over m\omega_x\right)^{1/2}
                 \left(\hbar\over m\omega_y\right)^{1/2}
                 \left(\hbar\over m\omega_z\right)^{1/2}
\end {equation}
is
the volume scale of the ground-state wave function.
(See also ref.\ \cite{review} for a review.)
So the appropriate infinite volume limit is
$\omega_x,\omega_y,\omega_z \to 0$ with $N\omega_x\omega_y\omega_z$ held fixed.

At low energies, the relevant measure of the strength of interactions is
the 2-body scattering length $a$.
We will assume that interactions are repulsive ($a>0$).
We will study the transition temperature $\Tc$ for Bose-Einstein condensation
of a dilute single-species gas as a function of the
total number of particles $N$ in the trap, in the infinite volume limit
just discussed.
One might naively anticipate there to be an expansion of the form
\begin {equation}
   \Tc(N) = T_0(N) \left[ 1 + c_1 \,{ a\over l}
                      + c_2 \left(a\over l\right)^2 + \cdots \right] ,
\end {equation}
where $T_0$ is the ideal gas result and $l$ is some characteristic
length of the ideal gas system.  As we'll review below,
the appropriate length scale for a trapped Bose gas is
the typical inter-particle separation
$l \sim \bar n^{-1/3} \sim N^{-1/6}\aho$
at the center of the trap.
The coefficient $c_1$ of the expansion for $\Tc$ has
been known for several years \cite{FirstOrder}.
In this paper, we calculate the next correction.  As we'll
discuss, this is the furthest one can go in the expansion without more
information about interactions than just the scattering length.
We'll find that $c_2$ depends logarithmically on $a/l$:
the actual expansion is of the form
\begin {equation}
   \Tc(N) = T_0(N) \left[ 1 + c_1 \, {a\over l}
                      + \left( c_2' \ln{a\over l} + c_2''\right)
                        \left(a\over l\right)^2 + \cdots \right] ,
\label {eq:Tcx}
\end {equation}
and we shall calculate the constants $c_2'$ and $c_2''$.
(The appearance of a related logarithm for {\it uniform}\/ gases has
been qualitatively discussed in Ref.\ \cite{logs}.
For a calculation of the second-order relationship between
$\Tc$ and the central density $\bar n$ in an arbitrarily wide trap,
which is also the relationship $\Tc(n)$ for a uniform gas, see
Ref.\ \cite{boselog}.)

Some aspects of the Bose-Einstein condensation phase transitions are
perturbatively calculable, and others are not.
In a dilute Bose gas, the physics of fluctuations associated with relatively
short distance scales is perturbative, while that associated with
critical behavior on relatively long distance scales is not.
In the case of a uniform Bose gas (that is, a Bose gas in an infinite
square well potential rather than a harmonic potential), the first-order
shift in $\Tc$ is sensitive to critical fluctuations and so is
non-perturbative.  That shift has recently been calculated using lattice
simulations \cite{prokofev,boselat1,boselat2} and
has previously been estimated in a wide
variety of ways \cite{stoof,gruter,holzmann,krauth,baym1,baymN,arnoldN}.
In contrast, the first-order shift for
a gas trapped in a harmonic potential (parametrized by $c_1$)
{\it is} calculable using perturbation
theory \cite{FirstOrder}.  As we shall see, the second-order logarithmic
coefficient $c_2'$ is also calculable in perturbation theory, but
the constant $c_2''$ under the log is not.
We shall calculate $c_2''$ by relating it to measurements that have been
made in lattice simulations of the phase transition in three-dimensional
O(2) field theory \cite{boselat2}.

We should emphasize that expansions of physical quantities
in $a/l$ cease to correspond to
{\it perturbative}\/ expansions in $a/l$, once one reaches the
orders we have asserted are non-perturbative.
The failure of perturbation
theory in describing generic second-order transitions has been known for
decades.
This breakdown typically manifests in perturbation theory as the appearance
of infrared infinities in the coefficients of the
perturbative expansion beyond a certain order.

There is a simple way to relate the problem of a Bose gas in an
arbitrarily wide harmonic potential with that of a uniform Bose gas.
In the infinite volume limit $\omega_x,\omega_y,\omega_z \to 0$ of the
harmonic trap problem, the trapping potential becomes everywhere
{\it locally}\/ flat over any fixed distance scale (such as the
typical inter-particle spacing).  Locally, the problem can then be
treated as a uniform gas in the presence of a $\x$-independent potential,
and an $\x$-independent potential can be absorbed into a redefinition of
the chemical potential.  For example, if the original chemical potential
was $\bar\mu$, then the effective chemical potential at a position $\x$
is $\bar\mu - V(\x)$.  For arbitrarily wide traps,
the total number of particles in the system is then
related to chemical potential and temperature by
\begin {equation}
   N = \int d^3x \> n\Bigl(T, \, \bar\mu - V(\x)\Bigr) ,
\label{eq:N}
\end {equation}
where $n(T,\mu)$ is the uniform gas result for the number density at
a chemical potential $\mu$.

In a trap, the effective chemical potential
$\bar\mu - V(\x)$ is highest at the center, where $V(\x)=0$, and this is
where the condensate first forms as the system is cooled.%
\footnote{
   A reminder about signs: Recall that, for a uniform Bose gas,
   $\mu$ is negative at high temperature and increases (moves towards
   zero) as the system is cooled.
}
For a uniform gas, let $\muc(T)$ be the critical value $\mu$ of
the chemical potential at a given temperature $T$.  Then
(\ref{eq:N}) becomes
\begin {equation}
   N = \int d^3x \> n\Bigl(\Tc, \, \muc(\Tc) - V(\x)\Bigr) .
\label{eq:Nc}
\end {equation}
If we knew $n(T,\mu)$ and $\muc(T)$ for a uniform gas, we could
then use (\ref{eq:Nc}) to solve for $\Tc$ for a gas of $N$ particles
in an arbitrarily wide trap.

In the next section, we review in more detail the
physical scales of the problem and explain why, for the purposes of
applying (\ref{eq:Nc}) to second order, it is adequate to use perturbation
theory for the uniform gas result $n(T,\mu)$.  We also explain why
perturbation theory is inadequate to find the uniform gas result
$\muc(T)$ at second order.
The second-order perturbative result for $n(T,\mu)$ can be extracted
from the literature \cite{n1,n2},
and in Section \ref{sec:N} we step through the simple
exercise of applying that old result to determine the relation
(\ref{eq:Nc}) between $N$, $\muc$, and $\Tc$ at second order.
Then, in Section \ref{sec:muc}, we take on the less trivial step
of showing how the second-order value of $\muc(T)$ can be related
to existing results from lattice simulations of O(2) scalar field theory
in three dimensions.
We put everything together in Section \ref{sec:final},
giving our final answer for the second-order term of the
expansion (\ref{eq:Tcx}) of $\Tc$.
In Section \ref{sec:higher},
we discuss the nature of yet higher-order corrections and explain
why they require more knowledge of 2-body scattering than just the
scattering length $a$.
In Section \ref{sec:wide}, we briefly discuss parametrically
how wide a trap must be for our ``arbitrarily wide trap limit'' results
to be valid at second order.
Finally, we conclude in Section \ref{sec:conclusion} with a brief
example of how big the second-order effects might be in a particular
experimental situation.
Various details and diversions are saved for appendices, including a modern
field-theory rederivation and verification
of the old perturbative
result for $n(T,\mu)$ that we take from
Huang, Yang, and Luttinger \cite{n1,n2}.


\section {Scales and Effective Theories}

\subsection {The uniform gas}

Before proceeding to a Bose gas in a harmonic trapping potential,
let's first review the basic scales
relevant to the phase transition of a uniform gas.
The generic condition that the gas is dilute is that the
two-particle scattering length $a$ be small compared to the
typical inter-particle separation $l \sim n^{-1/3}$, where
$n$ is the number density.  The Bose-Einstein condensation phase
transition occurs when the typical de Broglie wavelength
\begin {equation}
   \lambda \equiv \hbar \sqrt{2\pi\beta/m},
\label {eq:lambda}
\end {equation}
becomes of order the
inter-particle separation $l$.  Then $a \ll \lambda \sim n^{-1/3}$.

At the phase transition, the interaction can be treated perturbatively
for analyzing short-distance physics but, as with most second-order
phase transitions, interactions cannot be treated perturbatively for
analyzing long-distance physics.  A distance scale that will be
of interest is the dividing line between these two regimes.
As we shall review below, this scale is
$\lambda^2/a \sim n^{-2/3}/a$.  At the transition, there is then a hierarchy
$\lambda^2/a \gg \lambda \gg a$ of physically relevant distance scales for
a dilute Bose gas.

We will now briefly review the description of the dilute Bose gas system
in terms of effective field theories, and we'll then turn to the
effective field theory description relevant to the long distance physics
at the critical point \cite{baym1}.
This will provide a clean way to
review the origin of the non-perturbative scale $\lambda^2/a$, and we will
need to make use of such effective theories later in our discussion
of the critical chemical potential $\muc(T)$ for a uniform gas.

\subsection {The action}

It is well known that, at distance scales large compared to the scattering
length $a$, an appropriate effective theory for a dilute Bose gas is the
second-quantized Schr\"odinger equation, together with a chemical potential
$\mu$ that couples to particle number density $\psi^*\psi$, and a
$|\psi|^4$ contact interaction that reproduces low-energy scattering
\cite {review}.
The corresponding Lagrangian is%
\begin {equation}
   {\cal L} = \psi^* \left(
        {i\hbar} \, \partial_t + {\hbar^2\over 2m} \, \nabla^2
        + \mu - V(\x) \right) \psi
    - {2\pi\hbar^2 a\over m} \, (\psi^* \psi)^2 .
\label {eq:L1}
\end {equation}
The identification of the coefficient of the $(\psi^*\psi)^2$ interaction
with $2\pi\hbar^2a/m$ is technically only valid at leading order in the
interaction strength but, as we'll review later, doesn't change at
second order if one uses dimensional regularization \cite{eric}.
We'll also later discuss (in section \ref{sec:higher})
the size of corrections to the  effective theory
due, for instance, to energy
dependence of the cross-section or 3-body interactions.
It will turn out that such corrections can be ignored for the
purpose of computing $\Tc$ to second order.

To study (\ref{eq:L1}) at finite temperature, apply the standard imaginary
time formalism, so that $t$ becomes $-i\tau$ and imaginary time $\tau$
is periodic with period $\hbar\beta = \hbar/\kB T$.
The imaginary-time action is then
\begin {equation}
   \SI = \int_0^{\hbar\beta} d\tau \int d^3x \left[
        \psi^* \left(
        \hbar \, \partial_\tau - {\hbar^2\over 2m} \, \nabla^2
        - \mu + V(\x) \right) \psi
    + {2\pi\hbar^2 a\over m} \, (\psi^* \psi)^2
    \right] .
\label {eq:SI}
\end {equation}
As usual,
the field $\psi$ can be decomposed into imaginary-time
frequency modes with Matsubara
frequencies $\omega_n = 2\pi n/\hbar\beta$.


\subsection {Non-perturbative physics in the uniform gas}
\label{sec:npscale}

We'll now specialize the preceding to the uniform gas case
$V(\x) = 0$ and will discuss the system at or close to the critical
point.
For distances large compared to the thermal wavelength
(\ref{eq:lambda})
and sufficiently near the
transition so that $|\mu| \ll T$, the non-zero Matsubara frequencies
decouple from the dynamics, leaving behind an effective theory of
only the zero-frequency modes $\psi_0$, with the action becoming
\begin {equation}
   \hbar^{-1} \SI \to
   \beta \int d^3x \> \left[ \psi_0^* \left(
        - {\hbar^2\over 2m} \, \nabla^2
        - \mu_\eff \right) \psi_0
    + {2\pi\hbar^2 a\over m} \, (\psi_0^* \psi_0)^2 \right] ,
\label {eq:L2}
\end {equation}
up to corrections that again,
as we will discuss later (in section \ref{sec:higher}),
do not affect $\Tc$ at second order.
Eq.\ (\ref{eq:L2}) can be interpreted, if
desired, as the $\beta H$ of a classical 3-dimensional field
theory.

Finally, it is convenient to rewrite
$\psi_0 = \phi \sqrt{4\pi}/\lambda$ so that
the effective action becomes a conventionally normalized
U(1) field theory of a complex field $\phi$:
\begin {equation}
   S = \int d^3x \> \left[ (\grad\phi)^*\cdot(\grad\phi) + r \phi^*\phi
             + {u\over 6} \, (\phi^*\phi)^2 \right] ,
\label{eq:O2}
\end {equation}
where
\begin {equation}
   u = {96 \pi^2 a \over \lambda^2} \,.
\label {eq:u}
\end {equation}
We will henceforth refer to this effective theory as the ``three-dimensional''
effective theory, while referring to the original imaginary time theory
(\ref{eq:SI}) as the ``3+1 dimensional theory'' (for three space plus one
time dimension).
By writing $\phi = (\phi_1+i\phi_2)/\sqrt{2}$, the three-dimensional
effective theory may equivalently
be interpreted as an O(2) theory of two real fields with interaction
$(u/4!) (\phi_1^2+\phi_2^2)^2$.
The relationship of $r$ to the chemical potential $\mu$ and
other parameters of the original theory is a little more subtle,
because the $\phi^*\phi$ interaction is associated with an ultraviolet (UV)
divergence of the three-dimensional theory
that has to be renormalized.  We will discuss this relationship in
detail when we analyze $\muc(T)$ in section \ref{sec:muc}.  For the moment,
these details are unimportant.

There will be a line in the $(\mu,T)$ plane that corresponds to the
Bose-Einstein condensation phase transition.  In the long-distance
effective theory (\ref{eq:O2}), that line will correspond to a line in
the $(r,u)$ plane.  If we think of this line as determining $r$ in terms of
$u$, then the only physical scale in the problem of studying this
effective theory at the transition is $u$.  By dimensional analysis,
the distance scale of non-perturbative physics is therefore
$1/u \sim \lambda^2/a$, as asserted earlier.

It will be useful to understand how far away from the transition
one needs to go, as measured by $\muc-\mu$ at $T = \Tc$, in order for
the physics on {\it all} scales to be perturbative.  This will happen
when the correlation length $\xi$ is small compared to the scale
$1/u \sim \lambda^2/a$ of non-perturbative physics.
We can determine this condition on $\xi$
with a perturbative analysis.  In fact, it is sufficient to
consider a simple Gaussian ({\it i.e.}\ tree-level) approximation, where
$\mu_\eff$ in the effective three-dimensional theory (\ref{eq:L2})
is naively taken to be $\mu$, corresponding to $r = - 2 m \mu/\hbar^2$ in the
rescaled effective theory (\ref{eq:O2}).
In Gaussian approximation, $\muc = 0$.
The correlation length,
in Gaussian approximation, is $\xi \sim r^{-1/2}$, and so the condition
$\xi \ll 1/u$ becomes
\begin {equation}
   \muc - \mu \gg {\hbar^2 u^2\over m} \sim {\hbar^2 a^2\over m \lambda^4} \,.
\end {equation}
(An equivalent condition was discussed in the original work \cite{FirstOrder}
on the first-order result for $\Tc$ in a trap.)
We'll see later, in our more thorough discussion of the relationship between
$r$ and $\mu$ in section \ref{sec:muc}, that corrections to the Gaussian
approximation do not change this conclusion.

Finally, note that, by dimensional analysis, the non-perturbative contribution
to the critical value of $r$ in the three-dimensional $O(2)$ effective theory
(\ref{eq:O2}) must be of order $u^2$.  The
Gaussian approximation's identification of
$r$ with $-2 m \mu/\hbar^2$
then suggests that the non-perturbative contribution
to the critical value $\muc$ is of order
$\hbar^2 u^2/m$, which is second order in $a$.
As we'll see in section \ref{sec:muc}, this conclusion is correct.
This is the reason that, in order to calculate $\Tc$ to second order,
we must account for non-perturbative
physics in the determination of $\muc$.


\subsection {Gas in a harmonic trap}
\label {sec:gasintrap}

Now we turn to reviewing
scales in a harmonic trap.
One of the main points of this exercise will be to determine the
size of the region (at the transition) where the physics is
non-perturbative, relative to the size of the trapped gas cloud as a whole.
This will allow us to determine to what order one can use perturbative
calculations to relate $N$, $T$, and $\mu$ via (\ref{eq:N}).

For simplicity, we'll assume in
this discussion that $\omega_x \sim \omega_y \sim \omega_z$.
The relevant distance scales for a dilute Bose gas in an arbitrarily wide
harmonic trap, at the transition, are summarized
in Table \ref{tab:scales} in ascending order.
Most of this is just review of
simple, standard results \cite{review}, except for the scales of
non-perturbative physics in a harmonic trap, which we haven't seen
clearly discussed before.

\begin{table}[t]
\begin {center}
\tabcolsep=8pt

\begin {tabular}{|c|c|c|c|}             \hline
  $a$
    & scattering length
    & $a$
    & $a$
\\ \hline
  $l \sim \bar n^{-1/3} \sim \lambda$
    & \parbox{7.5cm}{\centering{
          \vspace{3pt}
          inter-particle separation at trap center; \\
          thermal wavelength
          \vspace{3pt}
      }}
    & $N^{-1/6}\aho$
    & $l$
\\ \hline
  $1/u$
   & \parbox{7.5cm}{\centering{
          \vspace{3pt}
          smallest wavelength of non-perturbative \\
          fluctuations near center of trap
          \vspace{3pt}
     }}
   & $N^{-1/3} \aho^2 / a$
   & $l^2/a$
\\ \hline
  $\aho$
    & size of the ground state (condensate)
    & $\aho$
    & $N^{1/6} l$
\\ \hline
  $\Rnp$
    & size of non-perturbative region
    & $N^{1/3} a$
    & $N^{1/3} a$
\\ \hline
  $\Rcloud$
    & size of entire gas cloud
    & $N^{1/6} \aho$
    & $N^{1/3} l$
\\ \hline
\end {tabular}
\end {center}
\caption
    {%
    \label {tab:scales}
       Distance scales for a dilute Bose gas in an arbitrarily wide
       harmonic trap at the phase
       transition.
       The scales are ranked in ascending order.
       Entries should be interpreted as
       representing orders of magnitude (parametric dependence)
       and not
       as precise definitions and equalities.
       The first column gives our notation for each scale.
       The third column shows how the scales depend
       on the ``experimental'' parameters $a$, $m$, $N$, and
       $\omega$, where $\aho \equiv (\hbar/m\omega)^{1/2}$.
       The last column shows a simple rewriting
       that makes the ordering of scales clear, given that our
       assumed limits can be phrased as $a$ fixed; $a \ll l$ (diluteness);
       and $l$ fixed with $N \to \infty$ (arbitrarily wide trap).
    }%
\end{table}

First, let's review the size and density of the cloud of Bose particles
at the phase transition.
As we'll reproduce below,
{\it most}\/ of the particles in the trap are in
the classical regime, and we can use the classical equipartition theorem
to find the width $\Rcloud$ of the cloud:
$\half m \omega^2 x^2 \sim \half \kB T$ yields
$\Rcloud \sim (\beta m \omega^2)^{-1/2}$.
The central density of particles is then of order
$\bar n \sim N/\Rcloud^3 \sim N (\beta m \omega^2)^{3/2}$, and
the separation of particles at the center of the trap is of order
$l \sim \bar n^{-1/3} \sim N^{-1/3} (\beta m \omega^2)^{-1/2}$.
The phase transition occurs when this separation is of order the
thermal wavelength (\ref{eq:lambda}),
giving $\kB T \sim N^{1/3} \hbar \omega$, and so
$l \sim \bar n^{-1/3} \sim \lambda \sim N^{-1/6}\aho$, as claimed
in Table\ \ref{tab:scales}.
The fact that $\kB T \sim N^{1/3} \hbar\omega \gg \hbar\omega$ in our
wide trap limit (which has $N \to \infty$) justifies the previous assertion
that, at the phase transition, most particles in the cloud can be treated
classically.

Now let's analyze the size of the region in which physics is non-perturbative
at the transition.  In our review of the uniform gas, we saw that
physics becomes completely perturbative when
$\muc-\mu \gg \hbar^2 a^2/m\lambda^4$.
In an arbitrarily wide trap, the effective value of $\mu$ is
$\bar\mu - {1\over2} m \omega^2 x^2$.  The condition for the existence of
non-perturbative physics at the transition is then
${1\over2} m \omega^2 x^2 \lesssim \hbar^2 a^2/m \lambda^4$, and the width of
the non-perturbative region is
$\Rnp \sim \hbar a/m\omega\lambda^2 \sim N^{1/3} a$.
Note that, even within this ``non-perturbative'' region, fluctuations with
small wavelengths ($\ll 1/u$) are still perturbative.

The relative volume of the non-perturbative region to the volume of the
entire gas cloud is $(\Rnp/\Rcloud)^3 \sim (a/l)^3$.
This means that non-perturbative contributions to the relation
$N = \int d^3x \> n(T, \, \bar\mu-V(\x))$ between $N$, $T$, and $\bar\mu$
are suppressed by more than
three powers of $(a/l)^3$.  It's {\it more}\/ than three powers because,
even in the relatively small
non-perturbative regime, the dominant contribution to the density comes
from typical particles, whose wavelengths are of order the thermal wavelength
$\lambda \gg 1/u$ and which can be treated perturbatively.
This makes the total suppression $(a/l)^4$.
In any case,
the conclusion is that there is no obstacle at second order in $a/l$ to using perturbation theory to derive the relationship between
$N$, $T$, and $\bar\mu$.


\section {\boldmath$\mbox{\lowercase{$n$}}(T,\mu)$ for a uniform gas
          and its application}
\label {sec:N}

\begin {figure}
\vbox{
   \begin {center}
      \epsfig{file=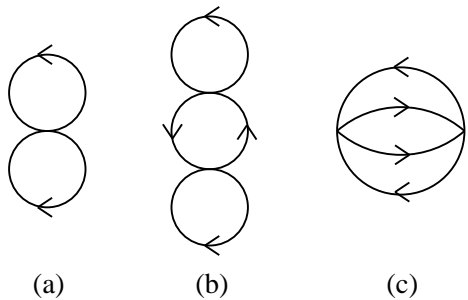,scale=.4}
   \end {center}
   \caption{
       Perturbative Feynman diagrams
       contributing to the pressure P at (a) first order and
       (b,c) second order.  Diagram (b)
       corresponds to the first $(a/\lambda)^2$ term in (\ref{eq:P})
       and diagram (c) to the second.
       \label{fig:P}
   }
}
\end {figure}

The second-order perturbative result for $n(T,\mu)$ can be easily extracted
from an old second-order result of Huang, Yang, and Luttinger \cite{n1,n2}
for the pressure of a uniform hard sphere gas:
\begin {eqnarray}
   P &=& {\kB T \over \lambda^3} \Biggl\{
        \Li_{5/2}(z)
        - {2a\over\lambda} [\Li_{3/2}(z)]^2
\nonumber\\ && \qquad
        + 8 \left(a\over\lambda\right)^2 \left(
              [\Li_{3/2}(z)]^2 \Li_{1/2}(z)
              + \sum_{i=1}^\infty \sum_{j=1}^\infty \sum_{k=1}^\infty
                  {z^{i+j+k}\over (i+k)(j+k)(ijk)^{1/2}}
          \right)
        + \cdots \Biggr\} ,
\label {eq:P}
\end {eqnarray}
where
\begin {equation}
   z = e^{\beta\mu}
\end {equation}
is the fugacity.
$\Li_n$ is the polylogarithm function, which for our purposes can be
considered as defined in terms of its series representation,
\begin {equation}
   \Li_n(z) = \sum_{s=1}^\infty {z^s\over s^n} .
\end {equation}
We have independently rederived and verified this result for the pressure.
For the sake of
any readers who might find a derivation in the language of field theory
[based on the Lagrangian (\ref{eq:L1})] a useful supplement to the original,
we give the derivation in Appendix \ref{app:P}. 
In the language of the imaginary-time field theory (\ref{eq:SI}), the
perturbative Feynman diagrams which correspond to the first- and second-order
terms in the pressure (\ref{eq:P}) are shown in Fig.~\ref{fig:P}.

\begin {figure}
\vbox{
   \begin {center}
      \epsfig{file=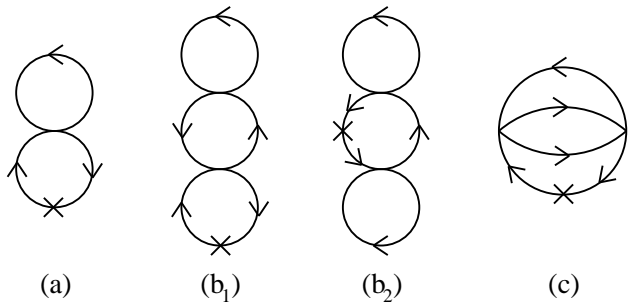,scale=.4}
   \end {center}
   \caption{
       Perturbative Feynman diagrams contributing to the number density
       $n = \langle \psi^* \psi \rangle$ at (a) first order and
       (b$_1$,b$_2$,c) second order.
       The cross corresponds to an insertion of the operator $\psi^* \psi$,
       whose expectation is being taken.
       \label{fig:n}
   }
}
\end {figure}

We now obtain $n$ as $\partial P/\partial\mu$:
\begin {eqnarray}
   n &=& {1 \over \lambda^3} \Biggl\{
        \sum_i {z^i\over i^{3/2}}
        - {2a\over\lambda} \sum_{ij}
                  {(i+j) z^{i+j}\over (ij)^{3/2}}
\nonumber\\ && \qquad
        + 8 \left(a\over\lambda\right)^2
          \sum_{ijk}
          (i+j+k) z^{i+j+k} \left[
             {1\over (ij)^{3/2} k^{1/2}}
             + {1\over (i+k)(j+k)(ijk)^{1/2}}
          \right]
        + \cdots \Biggr\} .
\nonumber\\
\label {eq:nresult}
\end {eqnarray}
Here and henceforth, indices of sums ($i$,$j$,$k$) implicitly
run from 1 to infinity.
(Most of the terms above could be written in terms of polylogarithms, but the
form shown is more convenient for the next step.)
We emphasize that this is a perturbative expansion and is valid only
in contexts where perturbation theory is applicable.%
\footnote{
   The same expansion was incorrectly applied in Ref.\ \cite{huangsilliness}
   to the problem of the {\it first}-order correction to $\Tc$ for a
   {\it uniform}\/ gas---a problem where perturbation theory breaks down.
}
In field theory language, the above result for $n$ corresponds to the
diagrams of Fig.\ \ref{fig:n}.

Now, use $\mu = \bar\mu - V(\x)$ and integrate over $\x$, as in
(\ref{eq:N}).  For the harmonic potential (\ref{eq:V}), the integrals
are simple Gaussian integrals, giving
\begin {eqnarray}
   N &=& \left( \kB T \over \hbar\omegaho \right)^3 \Biggl\{
        \sum_i {\bar z^i\over i^3}
        - {2a\over\lambda} \sum_{ij}
                  {\bar z^{i+j}\over (ij)^{3/2} (i+j)^{1/2}}
\nonumber\\ && \qquad
        + 8 \left(a\over\lambda\right)^2
          \sum_{ijk}
          {\bar z^{i+j+k} \over (i+j+k)^{1/2}} \left[
             {1\over (ij)^{3/2} k^{1/2}}
             + {1\over (i+k)(j+k) (ijk)^{1/2}}
          \right]
        + \cdots \Biggr\} ,
\nonumber\\
\label {eq:Nx}
\end {eqnarray}
where $\bar z \equiv e^{\beta\bar\mu}$ and
\begin {equation}
   \omegaho \equiv (\omega_x \omega_y \omega_z)^{1/3}
\end {equation}
is the geometric mean of the trap frequencies.

In Section \ref{sec:muc}, we will discuss the expansion
\begin {equation}
   \muc = \left[\muc^{(1)}{a\over\lambda}
             + \muc^{(2)}\left(a\over\lambda\right)^2 + \cdots \right] \kB T
\label {eq:mucx}
\end {equation}
of the critical value of $\bar\mu$
in powers of $a$.  The ideal gas result is $\muc = 0$.
Recall from previous discussion that the
second order term $\muc^{(2)}$ is not perturbatively calculable.%
\footnote{
   For those readers who like to think in terms of Feynman diagrams,
   there is a hand-waving, heuristic argument to see why there is a
   non-perturbative ${\cal O}(a^2)$ contribution to $\muc$.  The first-order
   contribution, given by the first diagram of Fig.\ \ref{fig:diverge},
   contributes ${\cal O}(a)$ to $\mu$.  Since this diagram is momentum
   independent, it can be absorbed into a renormalization of $\mu$, so that one
   need not consider higher-order diagrams that contain it as a sub-diagram
   (a point relevant to the discussion of divergences at higher
   orders).
   At the transition,
   the second diagram of Fig.\ \ref{fig:diverge}
   contributes ${\cal O}(a^2)$ times a logarithmic IR divergence,
   arising from the contribution where both loop frequencies are zero and
   the two loop momenta simultaneously approach zero.
   Three-loop contributions turn out to produce ${\cal O}(a^3)$ times linear
   IR divergences, four loops produce quadratic divergences, and so forth.
   Suppose we heuristically cut off these infrared divergences at a
   momentum scale $\Lambda_\IR$.  The perturbation series then turns out
   to look like
   $\beta\muc = O(a/\lambda) + O(a^2/\lambda^2) + O(a^3/\lambda^4\Lambda_\IR)
   + O(a^4/\lambda^6\Lambda_\IR^2) + \cdots$, where we have ignored logarithms
   such as $\ln(\lambda\Lambda_\IR)$.
   Imagine starting $\Lambda_\IR$ at some high value and then lowering it.
   The usefulness of the perturbative expansion will break down once
   we get to $\Lambda_\IR \sim a/\lambda^2$, which is just the non-perturbative
   scale $u$ discussed earlier.  For this $\Lambda_\IR$, all the terms in
   the series after the first become the same order, which is
   $O(a^2/\lambda^2)$.  This suggests that this is the order of
   a non-perturbative contribution to $\beta\muc$.  For a clean argument,
   however, one should instead refer to the analysis in the text.
}
We will later see that the second-order coefficient
$\muc^{(2)}$ contains a logarithm,
\begin {equation}
   \muc^{(2)} = A \ln{a\over\lambda} + B ,
\end {equation}
and that the coefficient $A$ of the logarithm is perturbatively calculable,
while $B$ is not.

In any case, we would like to insert the expansion (\ref{eq:mucx})
for $\muc$ into the expansion (\ref{eq:Nx}) for $N$.
The problem is a little more complicated than simply Taylor series expanding
the individual terms of the sums of (\ref{eq:Nx}) in $\bar\mu$, because
such a procedure would lead to unregulated infrared
logarithmic divergences at second order.
We derive the small $\bar\mu$ expansion of (\ref{eq:Nx})
in detail in Appendix \ref{app:Nx}, with the result
\begin {eqnarray}
   N &=& \left(\kB T \over \hbar\omegaho\right)^3 \Biggl\{
     \zeta(3)
     + {a\over\lambda}\Biggl[
         \zeta(2) \, \bar\mu^{(1)}
         - 2 \sum_{ij} {1\over i^{3/2} j^{3/2} (i+j)^{1/2}}
       \Biggr]
\nonumber\\ && \qquad
     + \left(a\over\lambda\right)^2 \Biggl[
         {\textstyle{3\over4}} [\bar\mu^{(1)}]^2
         + \zeta(2) \, \bar\mu^{(2)}
         - 2 \, \bar\mu^{(1)}
              \sum_{ij} {(i+j)^{1/2}-i^{1/2}-j^{1/2}\over i^{3/2}j^{3/2}}
         - 4 \, \zeta(\threehalf) \, \bar\mu^{(1)}
\nonumber\\ && \qquad\qquad
         + 8 \sum_{ijk} {1\over (ij)^{3/2} k^{1/2}} \left(
              {1 \over (i+j+k)^{1/2}} + {ij\over(i+k)(j+k)(i+j+k)^{1/2}}
              - {1\over k^{1/2}}
           \right)
       \Biggr]
\nonumber\\ && \qquad
     - {1\over2} \left(a\over\lambda\right)^2
       \left(\bar\mu^{(1)} - 4 \zeta(\threehalf)\right)^2
       \ln\left(-{\bar\mu^{(1)}a \over\lambda}\right)
     + O\!\left(a\over\lambda\right)^3
   \Biggr\} .
\label{eq:Nx2}
\end {eqnarray}

The logarithmic term at the end is the manifestation of the infrared
logarithm just mentioned.  In fact, at the critical point, the coefficient
of this logarithm vanishes because
\begin {equation}
   \muc^{(1)} = 4\,\zeta(\threehalf) .
\label {eq:muc1}
\end {equation}
A diagrammatic interpretation of why the logarithm vanishes
is given at the end
of Appendix \ref{app:Nx}.
The first-order result (\ref{eq:muc1}) for $\muc^{(1)}$ can be
derived using mean field theory, and a
discussion in the context of trapped Bose gases may be found in
the original first-order derivation of $\Tc$ \cite{FirstOrder}.  We
will rederive it in the next section, along with the second-order
coefficient $\muc^{(2)}$.  For the moment, though, let's use the
known first-order result (\ref{eq:muc1}) to solve for $\Tc$ in terms
of $\muc^{(2)}$.  Inverting (\ref{eq:Nx2}) gives
\begin {mathletters}%
\label {eq:Tcxmu2}%
\begin {equation}
   \Tc = T_0 \left[ 1 + c_1 \, {a\over\lambda_0}
             + c_2 \left(a\over\lambda_0\right)^2
             + O\!\left(a\over\lambda_0\right)^3 \right] ,
\end {equation}%
\begin {eqnarray}
   c_1 &=& {2\over 3\,\zeta(3)} \left[
            \sum_{ij} {1\over i^{3/2} j^{3/2} (i+j)^{1/2}}
            - 2\zeta(2)\zeta(\threehalf) 
         \right]
       \simeq -3.426~032 ,
\label {eq:c1}
\\
   c_2 &=& C_2 - {\zeta(2)\over3\,\zeta(3)} \, \muc^{(2)} ,
\end {eqnarray}%
\end {mathletters}%
\begin {eqnarray}
   C_2 &=&
      \fivehalf \, c_1^2
      + {4\over3\,\zeta(3)} \Biggl[
          \zeta(\threehalf)^2
          + 2 \zeta(\threehalf)
              \sum_{ij} {(i+j)^{1/2}-i^{1/2}-j^{1/2}\over i^{3/2}j^{3/2}}
\nonumber\\ && \qquad\qquad
         - 2 \sum_{ijk} {1\over (ij)^{3/2} k^{1/2}} \left(
              {1 \over (i+j+k)^{1/2}} + {ij\over(i+k)(j+k)(i+j+k)^{1/2}}
              - {1\over k^{1/2}}
           \right)
       \Biggr]
\nonumber\\
   &\simeq& 21.4,
\label {eq:C2}
\end {eqnarray}
where
\begin {equation}
   T_0 = \left(N \over \zeta(3)\right)^{1/3} {\hbar\omegaho\over \kB}
\end {equation}
is the ideal gas result and
\begin {equation}
   \lambda_0 = \sqrt{2\pi\hbar^2 \over m\kB T_0}
    = \sqrt{2\pi} \left(N \over \zeta(3)\right)^{-1/6} \aho
\end {equation}
is the corresponding thermal wavelength.
The first-order result is the same as that found in Ref.\ \cite{FirstOrder}.%
\footnote{
   The sum in (\ref{eq:c1}) is expressed in a slightly different
   form than in Ref.\ \cite{FirstOrder}.  The relation is that
  \[
     \sum_{ij} {1\over i^{3/2} j^{3/2} (i+j)^{1/2}}
     = \sum_{ij} {(i+j)\over i^{3/2} j^{3/2} (i+j)^{3/2}}
     = 2 \sum_{ij} {1\over i^{3/2} j^{1/2} (i+j)^{3/2}} ,
  \]
  where we've used $i \leftrightarrow j$ in the last step.
}
Results for the individual sums appearing above are listed in
Appendix \ref{app:sums}.


\section {\boldmath$\muc(T)$ for a uniform gas}
\label {sec:muc}

\subsection {Overview}

We'll now address how to relate the chemical potential $\mu$ appearing in
the original 3+1 dimensional theory (\ref{eq:L1}) to the
parameter $r$ of the effective
three-dimensional theory (\ref{eq:O2}).  The critical value $r_\c$ of
$r$ can be extracted from lattice simulations of the latter theory
\cite{boselat2}, which will then allow us to determine the critical value
$\muc(T)$ of $\mu$.

Effective theories, such as the three-dimensional O(2) model,
have long been used to
describe long-distance physics at second-order phase transitions.
Such use of effective theories is often restricted to studies of
universal quantities, such as critical exponents, because the relationship
between the parameters of the effective theory and a more fundamental
description of the system cannot be computed systematically.
The situation is quite different for dilute Bose gases near the
phase transition: the short distance
scale $\lambda$ at which the long-distance three-dimensional effective theory
description (\ref{eq:O2}) breaks down is a scale at which the physics is
{\it perturbative}\/ (since $\lambda \ll 1/u$).  One may therefore
perform a perturbative calculation to relate $r$ to $\mu$, even though
the long-distance physics at the transition is non-perturbative.

Such perturbative matching of the parameters of effective theories with
underlying physics has a long history in field theory.
It has been applied in a number of problems, including
lattice field theory \cite{symanzik},
Bose condensates at zero temperature \cite{Bose},
relativistic corrections to non-relativistic QED \cite{QED},
heavy quark physics \cite{heavy quarks},
ultrarelativistic plasmas \cite {Braaten&Nieto},
and non-relativistic plasma physics \cite{nonrel plasma}.
For a general discussion, see also Ref. \cite{effective}.
The basic idea is to formally compute, in perturbation theory,
some number of infrared physical quantities in both the effective theory
and the more fundamental theory.  By equating the results from the
two theories, one can then solve for the parameters of the
effective theory (to the order desired).

The perturbative computations are performed using any convenient
infrared regulator (though it must be the same regulator in both
theories).  The perturbation series for various physical
quantities will be badly behaved if one removes the infrared regulator
since, in our case at least, the infrared physics is non-perturbative.
But this bad infrared behavior will cancel out in the perturbation series
derived for the parameters of the effective theory, and so one may
safely remove the infrared regulator at the end of the matching
calculation.  This is a reflection of the fact that the difference
between the effective theory and underlying theory has to do with
short-distance physics, and short-distance physics is perturbative
(in the cases where perturbative matching is applicable).

The relevant distance scale of physics for the matching calculation is
the short-distance scale $\lambda$ where the three-dimensional O(2)
theory breaks down.  The corresponding energy scale is therefore of order
$\hbar^2 / m \lambda^2 \sim \kB T$, which is simply the
typical energy of particles in the gas.
This scale is large compared to the size of
the chemical potential at the transition (\ref{eq:mucx}), which is
of order $(a/\lambda) \kB T$.  Therefore, for the purpose of doing
a matching calculation, the chemical potential $\mu$ may be treated
as a perturbation.  In combination with the use of dimensional
regularization, this turns out to be very convenient computationally.

With $\mu$ treated perturbatively, the imaginary-time Feynman rules
for the original 3+1 dimensional action (\ref{eq:SI}) are shown, for reference,
in Table \ref{tab:feyn}.
The analogous rules for the three-dimensional O(2) effective theory are also
shown.
When discussing the evaluation of Feynman diagrams,
we will always set $\hbar=\kB=1$ in order to avoid cluttering up equations and
discussions of conventions.
We've specialized to the case of a uniform
gas by taking $V(\x)=0$, and we've introduced the shorthand
\begin {equation}
       \omega_k \equiv {k^2\over 2m} \,.
\end {equation}
We will use the notation $k_0$, $l_0$, $p_0$, ... to designate the
Matsubara (imaginary time) frequencies associated with propagators with
momenta $\k$, $\l$, $\p$, ....

\begin{table}[t]
\begin {center}
\tabcolsep=8pt

\begin {tabular}{|c|c|c|}             \hline
  & 3+1 dim.\ theory of $\psi$
  & 3 dim.\ theory of $\phi$
\\ \hline
  \centering{$\vcenter{\hbox{\epsfig{file=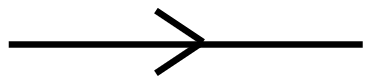,scale=.3}}}$}
  & $\displaystyle{1\over i k_0 + \omega_\k}$
  & $\displaystyle{1 \over k^2}$
\\
  \centering{$\vcenter{\hbox{\epsfig{file=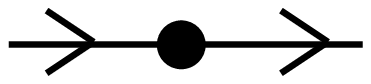,scale=.3}}}$}
  & $+\mu$
  & $-r$
\\
  \centering{$\vcenter{\hbox{\epsfig{file=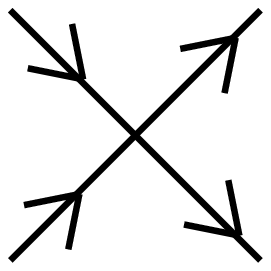,scale=.3}}}$}
  & $\displaystyle{-{8\pi a\over m}}$
  & $\displaystyle{-{2u\over3}}$
\\ \hline
\end {tabular}
\end {center}
\caption
    {%
    \label {tab:feyn}
       Feynman rules, appropriate for a matching calculation in a uniform Bose
       gas, for the original 3+1 dimensional theory (\ref{eq:SI}) of $\psi$
       and the effective three-dimensional theory (\ref{eq:O2}) of $\phi$.
       We have set $\hbar=\kB=1$.
       The variable $k_0$ represent the Matsubara frequency of the field, while
       $\omega_k \equiv k^2/2m$.
       At finite temperature, loop frequencies $l_0$ are summed over the
       discrete values $l_0 = 2\pi n T$ with $n$ any integer.
       In dimensional regularization with the \MSbar\ renormalization scheme,
       a factor of
       $\ren^\eps = (e^{\gammaE/2}\bar\ren/\sqrt{4\pi})^\eps$ should also be
       associated with each 4-point vertex but has not been explicitly shown
       above.
    }%
\end{table}

In our case, the short-distance length
scale $\Lambda^{-1}$ at which the three-dimensional theory breaks down is
of order $n^{-1/3} \sim \lambda$, as we've discussed before.
In principal, a long-distance effective theory can correctly describe
physics at an infrared wavelength scale $k \ll \Lambda$ to any desired
order in $k/\Lambda$.  However, as one pushes the description to higher
and higher powers of $k/\Lambda$, one must add more and more corrections
to the action of the effective theory, in the form of interactions that
are more and more infrared irrelevant (in the sense of the renormalization
group)---that is, interactions with higher scaling dimension.
In our case, the long-distance physics scale of interest is the
non-perturbative scale $1/u$, and powers of $k/\Lambda$ translate into
powers of our expansion parameter $u \lambda \sim a/\lambda \ll 1$.
We shall discuss later why including such corrections, such as a
$|\phi^*\grad\phi|^2$ terms in the effective Lagrangian,
would in particular not affect $\Tc$ at second order.
[We will also give a similar discussion of $(\phi^*\phi)^3$, which is
a marginal operator in three dimensions.]
For now, though, we shall simply
ignore the issue and push ahead with the matching calculation.

The action for a given effective theory can be written in a variety of
equivalent ways by making field redefinitions, such as $\phi \to c\phi$
or $\phi \to (1 + \epsilon \nabla^2 + \cdots) \phi$, etc.  Our convention shall
be to insist that the fields of the three-dimensional and
3+1 dimensional theories match up, to whatever order in $k/\Lambda$ we
are working, as
\begin {equation}
   \psi(0,\k) = {\sqrt{2mT}\over \hbar} \, \phi(\k) .
\end {equation}
This was our definition of $\phi$ in the more cavalier discussion in the
introduction.  The frequency $k_0$ of $\psi(k_0,\k)$ denotes imaginary-time
frequency.  So one of our matching conditions will be that the
inverse Green functions match up as
\begin {equation}
   G^{-1}_\psi(0,\k) = {G^{-1}_\phi(\k) \over 2m}
\label {eq:Gequate}
\end {equation}
to the relevant order in $k/\Lambda$.
In the presence of interaction, this definition of $\phi$ might fix the
normalization of the $(\grad\phi)^*(\grad\phi)$ term in the action
(\ref{eq:O2}) to be different from 1.
Our three-dimensional effective theory should therefore be
written in the somewhat more general form
\begin {equation}
   S = \int d^{3-\eps}x \> \left[ Z_\phi (\grad\phi)^*\cdot(\grad\phi)
             + r_\eff \phi^*\phi
             + {u_\eff\over 6} (\phi^* \phi)^2
             + \mbox{(higher-dimensional operators)}
       \right] ,
\end {equation}
where $Z_\phi$ can deviate from one beyond leading order.
In principle, we need
to determine the parameters $Z_\phi$, $r_\eff$, and $u_\eff$ (and any
higher-dimensional operators, if they were required at a desired order)
by matching.


\subsection {UV regularization}
\label{sec:UV}

\begin {figure}
\vbox{
   \begin {center}
      \epsfig{file=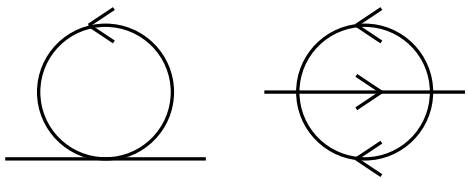,scale=.7}
   \end {center}
   \caption{
       Fundamental UV-divergent diagrams in the three-dimensional theory
       (\ref{eq:O2}).  Dimensional regularization automatically removes
       the linear divergence of the first diagram and regulates the
       logarithmic divergence of the second diagram as $1/\eps$.
       \label{fig:diverge}
   }
}
\end {figure}

Before starting on matching, we must first unambiguously
define the parameters of
our theories.  The three-dimensional long-distance O(2) theory (\ref{eq:O2})
is super-renormalizeable,
but there are UV divergences associated with the $\phi^*\phi$ interaction.
Diagrammatically, these divergences are associated with the graphs
of Fig.\ \ref{fig:diverge}.
In order to give the coefficient $r$ a well-defined meaning, we need to
specify a regularization and renormalization scheme.
By far the most convenient regularization scheme for perturbative matching
calculations is dimensional regularization.  We shall replace the
number $d=3$ of spatial dimensions by $d=3-\eps$, taking $\eps \to 0$ at
the end of the day.

To define a finite, renormalized value of $r$,
we will use the modified
minimal subtraction (\MSbar) scheme
with a renormalization scale $\bar\ren$.
The theory is then
\begin {equation}
   S = \int d^{3-\eps}x \> \left[
             Z_\phi (\grad\phi)^*\cdot(\grad\phi)
             + r_\bare \phi^*\phi
             + \ren^\eps
                     {u_\eff\over 6} \, (\phi^*\phi)^2 \right] ,
\label {eq:O2ren}
\end {equation}
with the relation
\begin {equation}
   r_\bare = \rms + {1\over (4\pi)^2\eps}\left(u\over3\right)^2
\label{eq:rMSbar}
\end {equation}
between the bare coupling $r_\bare$ and the renormalized
coupling $\rms(\bar\ren)$, and where
\begin {equation}
   \ren \equiv {e^{\gammaE/2}\over\sqrt{4\pi}} \, \bar\ren .
\label{eq:MSbar}
\end{equation}
[The factor of $e^{\gammaE/2}/\sqrt{4\pi}$
in (\ref{eq:MSbar}) is what distinguishes
modified minimal subtraction (\MSbar) from unmodified minimal subtraction
(MS); the difference between the two schemes amounts to nothing more
than a multiplicative redefinition of the renormalization scale.]

\begin {figure}
\vbox{
   \begin {center}
      \epsfig{file=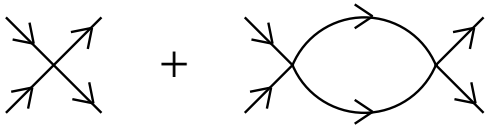,scale=.5}
   \end {center}
   \caption{
       The 2-particle scattering amplitude at second-order in the interaction
       strength.  (Note that there is no t-channel analog of the second
       diagram, since such a diagram vanishes in a non-relativistic theory.)
       \label{fig:4point}
   }
}
\end {figure}

The original 3+1 dimensional effective theory (\ref{eq:L1}) is
not renormalizeable and also requires UV regularization, and we will
again use dimensional regularization.  At second order
in the interaction strength (the order relevant to our calculation),
the only UV divergence is a well-known linear divergence
associated with the second
diagram of Fig.\ \ref{fig:4point}, which can be absorbed
into a redefinition of the coefficient of the $(\psi^*\psi)^2$ interaction.
To relate this coefficient to the physical scattering length $a$, one needs to
regularize the theory and then compute the zero-energy limit $\sigma(0)$
of the 2-particle cross-section (at zero temperature and density),
since $a$ is defined by $8\pi a^2 \equiv \sigma(0)$ for identical particles.
In dimensional regularization, however, the loop integral for the second
diagram in Fig.\ \ref{fig:4point} vanishes at zero temperature and density,
and so there is no second-order correction to $\sigma(0)$.
The coefficient of the quartic interaction therefore remains its
tree-level value $2\pi\hbar^2a/m$, as in (\ref{eq:L1}) \cite{eric}.

This property of dimensional regularization is a simple consequence of
dimensional analysis.
At zero energy (i.e.\ zero external momenta),
the second diagram in Fig.\ \ref{fig:4point} is proportional to the loop
integral
\begin {equation}
   \int {d l_0 \> d^{3-\eps} l
       \over(i l_0 + \omega_l) (-i l_0 + \omega_l)}
\end {equation}
All of the parameter dependence of this integral can be factored out
by rescaling $l_0$ by a factor of $2m$.  The rescaled integral
\begin {equation}
   \int {d l_0 \> d^{3-\eps} l
       \over(i l_0 + l^2) (-i l_0 + l^2)}
\end {equation}
has dimensions of (length)$^{-1+\eps}$ but no dimensional parameters to make
up that dimension.  The only dimensionally consistent answer is zero.
In most regularization schemes other than dimensional regularization, there
{\it are} still dimensionful parameters in the integral associated with
cut-off scales, and the integral would not be zero.  For example, if
we regulated with a UV cut-off $\Lambda$ on $l$, the integral would
give a non-zero result proportional to $\Lambda$ in $d=3$.

\subsection {Matching of \boldmath$Z_\phi$}

We want to calculate the critical value $\muc$ to next-to-leading order
[i.e.\ $\mu^{(2)}$ in the expansion (\ref{eq:mucx})].
One might
expect this to  require
knowing the parameters $Z_\phi$, $r_\eff$, and $u_\eff$ of the effective
theory to next-to-leading order.
In fact, as we shall see, dimensional regularization organizes the calculation
in such a way that we only need to compute $r_\eff$, which is three-dimensional
analog of $\mu$.  But let us briefly discuss the matching of $Z_\phi$
anyway, as a simple warm up.

The matching of $Z_\phi$ is trivial because
the first-order contribution to the inverse propagator, given by the
first diagram of Fig.\ \ref{fig:diverge}, does not have any momentum
dependence.  That is, equating the inverse
propagators of the two theories as in (\ref{eq:Gequate}) gives
\begin {equation}
   {k^2\over 2m} + \mbox{($\k$ independent)} + {\cal O}(a^2)
   = {1\over 2m}
       \left[ Z_\phi k^2 + \mbox{($\k$ independent)} + {\cal O}(a^2) \right] ,
\end {equation}
where ${\cal O}(a^2)$ indicates corrections that are formally second order in
perturbation theory.
So
\begin {equation}
   Z_\phi = 1 + {\cal O}(a^2) .
\end {equation}

In this paper, we will write $O(...)$ when displaying the full
parameter dependence of a correction (except possibly for logarithmic
factors) and write ${\cal O}(...)$ when just showing the dependence on
a particular parameter.
So $32 a^2/\lambda^2 = O(a^2/\lambda^2) = {\cal O}(a^2)$.
In matching calculations, where we are formally doing perturbation theory
with IR regularization,
${\cal O}(a^n)$ will just mean $n$-th order in perturbation theory.


\subsection {Matching of \boldmath$r$}

To match $r$, take the $\k=0$ case of matching (\ref{eq:Gequate})
the inverse Green functions:
\begin {equation}
   -\mu + \Pi_\psi(0,0) = {1 \over 2m} \left[ r_\bare + \Pi_\phi(0) \right] ,
\label {eq:rmatch}
\end {equation}
where $\Pi$ is the proper self-energy.

\begin {figure}
\vbox{
   \begin {center}
      \epsfig{file=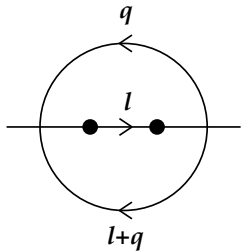,scale=.35}
   \end {center}
   \caption{
       An two-loop example of a diagram contributing to $-\Pi_\phi(0)$
       for a matching calculation.  $r$ is treated perturbatively.
       \label{fig:PIexample}
   }
}
\end {figure}

We will use dimensional
regularization to regulate the infrared divergences of perturbation
theory, as well as the UV divergences already discussed.
A well-known advantage of such use of dimensional regularization for
matching calculations is that then every loop diagram contributing to
$\Pi_\phi(0)$ vanishes by dimensional analysis arguments similar to the
one given in Section \ref{sec:UV}.  Consider, for example, the
contribution of Fig.\ \ref{fig:PIexample}.
In the three-dimensional effective theory,
this diagram is proportional to the loop integrals
\begin {equation}
   \int {d^{3-\eps}l \> d^{3-\eps}q \over l^6 q^2 |\l+\q|^2} ,
\end {equation}
which must vanish because there are no dimensionful parameters to make up
the dimensions of the result.  It is crucial here that there are no
external momenta, that $r$ may be treated as a perturbation for the purpose
of matching calculations, and that loop integrals are never dimensionless in
dimensional regularization.
The upshot is that the matching condition
(\ref{eq:rmatch}) becomes simply
\begin {equation}
   r_\bare = 2m \left[ -\mu + \Pi_\psi(0,0) \right] .
\label{eq:rmatch2}
\end {equation}

Another convenience of the vanishing, in dimensional regularization, of
loop diagrams in the three-dimensional theory is that we need not
keep track of the matching of $u$ and $Z_\phi$ if we're only interested
in the matching of $r$.  For instance, the one-loop contribution of the
first diagram in Fig.\ \ref{fig:diverge}
gives a contribution proportional to $u$
in three dimensions, and so a second-order calculation of $\Pi_\phi$
would require a second-order determination of $u$, if it weren't for the
fact that this diagram vanishes.

\begin {figure}
\vbox{
   \begin {center}
      \epsfig{file=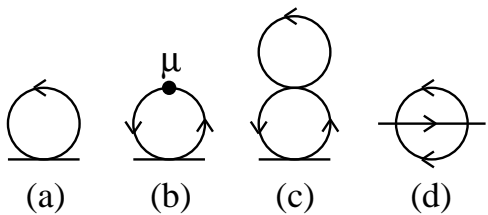,scale=.5}
   \end {center}
   \caption{
       Diagrams contributing to $-\Pi_\psi$.  $\mu$ has been treated
       perturbatively.
       \label{fig:r}
   }
}
\end {figure}

Returning to the 3+1 dimensional theory,
the diagrams which contribute to $\Pi_\psi$ up to second order are shown in
Fig.\ \ref{fig:r}, and all diagrams are to be evaluated at finite
temperature.  Diagram (a) gives the first-order contribution to
$\Pi_\psi$.  It gives
\begin {equation}
   \Pi^{\rm (a)}_\psi =
   {8\pi a\over m} \, \sumint_P {1\over i p_0 + \omega_p} ,
\end {equation}
where we introduce the shorthand notations
\begin {equation}
   \sumint_P \equiv T \sum_{p_0} \int_\p 
   \equiv T \sum_{p_0} \, \ren^\eps \! \int {d^dp \over (2\pi)^d} \, ,
\label {eq:defsumint}
\end {equation}
where $d{=}3-\eps$ is the number of spatial dimensions.
The $p_0$ sum can be performed by standard contour tricks,%
\footnote{
   For example, see Section 25 of Ref.\ \cite{FetterWalecka}.
}
yielding
\begin {equation}
   \Pi^{\rm (a)}_\psi
   = {8\pi a\over m} \int_\p \left[n(\omega_\p)+\half\right]
\end {equation}
where $n(\omega)$ is the Bose distribution function
\begin {equation}
   n(\omega) \equiv {1\over e^{\beta\omega}-1} \,.
\end {equation}
The integral of a constant vanishes in dimensional regularization
(again by dimensional analysis), and
the integral of $n(\omega_\p)$ can be carried out in three dimensions
to yield
\begin {equation}
   \Pi^{\rm (a)}_\psi
   = {4 a T\over\lambda} \, \zeta(\threehalf) + {\cal O}(\eps) .
\end {equation}

\begin {figure}
\vbox{
   \begin {center}
      \epsfig{file=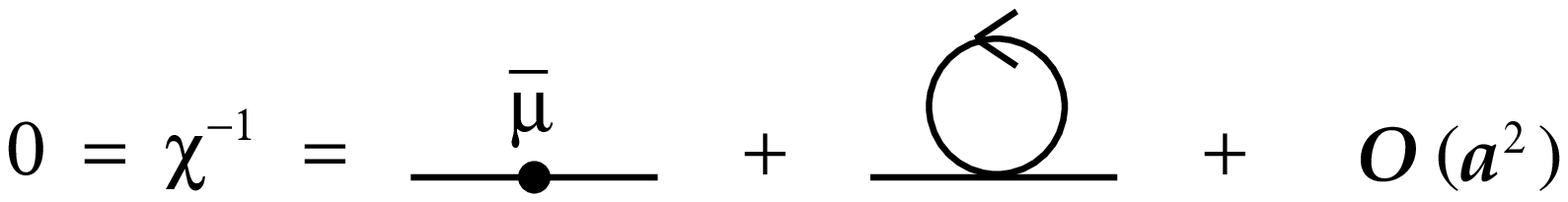,scale=.7}
   \end {center}
   \caption{
       The vanishing of the inverse susceptibility at the phase transition,
       expressed in terms of diagrams at first order in $a$.
       (Unlike Fig.\ \ref{fig:logs}, 
       these diagrams are evaluated at the center of the trap and not
       integrated over $\x$.)
       \label {fig:cancel}
   }
}
\end {figure}

At the critical point, diagrams (b) and (c) cancel each other and so need not
be computed individually.  The cancellation arises because
the inverse susceptibility $\chi_\psi^{-1} = -\mu + \Pi_\psi(0,0)$
will vanish at a second order phase transition.  This condition is shown
diagrammatically in Fig.\ \ref{fig:cancel}
at first order in perturbation theory.
As we've discussed in Section \ref{sec:npscale},
perturbation theory breaks down in
the calculation of $\mu$ at second order, but the first-order relation
of Fig.\ \ref{fig:cancel} is therefore reliable.%
\footnote{
  You may wonder why we've discussed the reliability of
  perturbation theory here when we've already asserted that perturbation
  theory is valid for the matching calculation.  The reason is that we're
  jumping ahead a little in order to streamline the calculation.
  The matching calculation can be done perturbatively because it
  involves only physics at the perturbative scale $\lambda$,
  but the subsequent
  solution for $r_\c$ (and therefore $\muc$) cannot, since it involves
  physics at the non-perturbative scale $1/u$.  Since we are using a
  result about $\muc$ to simplify our matching calculation at $\muc$,
  we need to be careful.
}
This relation implies that
diagrams (b) and (c) cancel at second order:
\begin {equation}
   \Pi_\psi^{\rm(b)} + \Pi_\psi^{\rm(c)} = {\cal O}(a^3)
   \quad
   \mbox{at $\mu=\muc$.}
\end {equation}

\subsubsection*{The sunset diagram}

\begin {figure}
\vbox{
   \begin {center}
      \epsfig{file=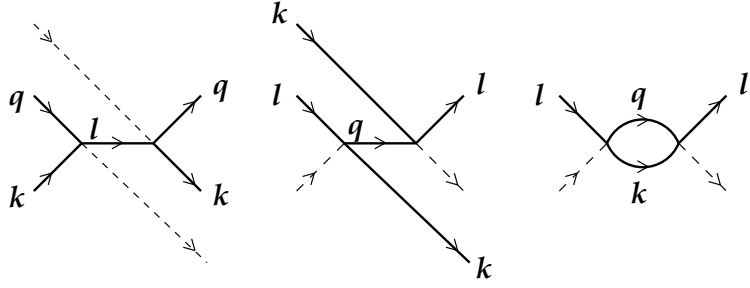,scale=.5}
   \end {center}
   \caption{
       Time ordered perturbation theory diagrams corresponding to
       the three terms of (\ref{eq:Pidn}).  Time flows from left to right,
       and the dashed line represents the zero external momentum (and
       zero frequency) in diagram (d) for $-\Pi_\psi(0,0)$.
       Each incoming momentum (other than the zero one) is associated with
       a Bose distribution factor $n$.
       \label {fig:timeorder}
   }
}
\end {figure}

We now turn to diagram (d), the sunset diagram:
\begin {equation}
   \Pi_\psi^{\rm(d)}(0) =
      - {1\over2} \left(8\pi a\over m\right)^2
      \sumint_Q \sumint_K { 1 \over
         (i q_0 + \omega_q) (i k_0 + \omega_k) [i(q_0+k_0)+\omega_{\q+\k}] } .
\label {eq:Pid}
\end {equation}
We review in Appendix \ref{app:sunset} how the loop frequency sums can be
done with standard contour tricks, with the result
\begin {equation}
   \Pi_\psi^{\rm(d)}(0) =
      - {1\over2} \left(8\pi a\over m\right)^2
      \int_{\q\k\l} \PP
         {[n(\omega_q)\,n(\omega_k) - 2 n(\omega_k)\,n(\omega_l) - n(\omega_l)]
         \over \omega_l - \omega_q - \omega_k}
         \, (2\pi)^d \delta^{(d)}(\l-\q-\k) .
\label {eq:Pidn}
\end {equation}
The symbol $\PP$ indicates the principal value prescription
\begin {equation}
   \PP \, {1\over x} = \Re \, {1\over x + i0^\pm} \,,
\label {eq:PP}
\end {equation}
where $0^\pm$ is an infinitesimal.
This prescription removes the spurious divergence associated with
$\omega_l - \omega_q - \omega_k \to 0$ (but not also $l \sim q \sim k \to 0$),
which is an artifact of this form of writing $\Pi^{\rm(d)}(0)$.
(See Appendix \ref{app:sunset}.)
We won't bother to explicitly
write the $\PP$ in what follows.
We note, as a side remark, that,
in the language of time-ordered perturbation theory (in real time),
the three terms of
(\ref{eq:Pidn}) correspond to the three diagrams of Fig.\ \ref{fig:timeorder}.

It's easy to see that the first term of (\ref{eq:Pidn}) vanishes, because
it is proportional to
\begin {equation}
   \int_{\q\k} { n(\omega_q)\,n(\omega_k)
         \over \omega_{\q+\k} - \omega_q - \omega_k }
   \propto
   \int_{\q\k} { n(\omega_q)\,n(\omega_k)
         \over |\q+\k|^2 - q^2 - k^2 }
   =
   \int_{\q\k} { n(\omega_q)\,n(\omega_k)
         \over 2\q\cdot\k } \,,
\end {equation}
which vanishes by $\q \to -\q$ (for any reasonable choice of regularization
scheme).  In Appendix \ref{app:sunset}, we show that the last term of
(\ref{eq:Pidn}) vanishes
as $\epsilon \to 0$ in dimensional regularization.
Only the second term of (\ref{eq:Pidn}) remains, giving
\begin {equation}
   \Pi_\psi^{\rm(d)}(0) =
      \left(8\pi a\over m\right)^2
      \int_{\k\l}
         {n(\omega_k)\,n(\omega_l)
         \over \omega_l - \omega_{\k-\l} - \omega_k}
      + {\cal O}(\eps).
\label {eq:Pidn2}
\end {equation}


\subsubsection*{Subtracting divergences of the sunset diagram}

Because $n(\omega_p) \to (\beta\omega_p)^{-1} \propto p^{-2}$ as
$p\to0$, the above integral (\ref{eq:Pidn2})
has a logarithmic infrared divergence
associated with $k \sim l \to 0$.
Unfortunately, the full integral in (\ref{eq:Pidn2}) is too complicated
for us to do in arbitrary dimensions, which would be the most
straightforward way to apply our chosen regularization scheme,
dimensional regularization.  It's convenient to instead explicitly isolate
the divergent IR behavior by rewriting
\begin {equation}
   \Pi_\psi^{\rm(d)}(0) =
      \left(8\pi a\over m\right)^2 \left[
      \int_{\k\l}
         {n(\omega_k)\,n(\omega_l) - n_0(\omega_k)\,n_0(\omega_l)
         \over \omega_l - \omega_{\k-\l} - \omega_k}
      + \int_{\k\l}
         {n_0(\omega_k)\,n_0(\omega_l)
         \over \omega_l - \omega_{\k-\l} - \omega_k}
      + {\cal O}(\eps) \right],
\label {eq:dsubIR}
\end {equation}
where
\begin {equation}
   n_0(\omega) \equiv {1\over\beta\omega} \,.
\end {equation}
The second integral vanishes in dimensional regularization for the usual
reason: it is
proportional to
\begin {equation}
   \int_{\k\l} {1\over k^2 l^2 (l^2-|\k-\l|^2-k^2)} ,
\end {equation}
which contains no dimensionful parameter to make up its dimensions.  So
\begin {equation}
   \Pi_\psi^{\rm(d)}(0) =
      \left(8\pi a\over m\right)^2
      \int_{\k\l}
         {n(\omega_k)\,n(\omega_l) - n_0(\omega_k)\,n_0(\omega_l)
         \over \omega_l - \omega_{\k-\l} - \omega_k}
      + {\cal O}(\eps) .
\label {eq:Pidn3}
\end {equation}
The above integral is infrared convergent and, if it weren't for the fact
that we've now introduced a UV divergence associated with
$k \sim l \to \infty$, we would be able to set $d=3$ in that integral
and ignore regularization issues.

To continue, it is useful to understand another way to interpret the infrared
behavior represented by the last term of (\ref{eq:dsubIR}):
\begin {equation}
   I_0 = \left(8\pi a\over m\right)^2 \int_{\k\l} {
         n_0(\omega_k)\,n_0(\omega_l)
         \over \omega_l - \omega_{\k-\l} - \omega_k
       } ,
\label {eq:I0}
\end {equation}
As discussed
before,
infrared physics
is dominated, in imaginary time, by the zero-frequency mode of the
field $\psi$.  $I_0$ turns out
to be the $q_0=k_0=0$ piece of the original frequency sums (\ref{eq:Pid})
representing diagram (d).  A quick way to see this is to note that
the integrand in $I_0$ above is the
high temperature limit ($\beta\to0$) of the original integrand in
(\ref{eq:Pidn2}).  But, if one goes all the way back to the original
imaginary-time frequency sums (\ref{eq:Pid}), the
integrand there is proportional to
\begin {equation}
    \sum_{q_0} \sum_{k_0} { 1 \over
         (i q_0 + \omega_q) (i k_0 + \omega_k) [i(q_0+k_0)+\omega_{\q+\k}] } ,
\end {equation}
with $q_0$ and $k_0$ of the form $2\pi n T$.
Only the $q_0 = k_0 = 0$ piece survives in the $\beta \to 0$ limit of
this integrand, and this establishes the correspondence.

It is important to note that non-zero frequency modes {\it do} contribute
to diagram (d) even in the infinitely high temperature limit, because the
limit does not commute with the integration over spatial momenta $\q$
and $\k$.  However, in our analysis so far, we have not yet performed the
$\q$ and $\k$ integrations, and it is okay to take limits of integrands
to see the correspondence of $I_0$ with the $q_0=k_0=0$ piece of diagram (d).

The upshot is that the infrared piece $I_0$ that we isolated from
diagram (d) is proportional to the same diagram evaluated
in a purely three-dimensional theory:
\begin {equation}
   I_0 = 
      - {T^2\over2} \left(8\pi a\over m\right)^2
      \int_{\q\k} { 1 \over
         \omega_q \omega_k \omega_{\q+\k} }
   =
      - 4m^3 T^2 \left(8\pi a\over m\right)^2
      \int_{\q\k} { 1 \over
         q^2 k^2 |\q+\k|^2 }
   \,.
\label {eq:I0b}
\end {equation}
This diagram is logarithmically divergent in both the infrared and
ultraviolet, just as the original expression (\ref{eq:I0}) for $I_0$,
and it vanishes in dimensional regularization.  The UV divergence of
our current expression (\ref{eq:Pidn3}) for diagram (d) came from the
UV divergence of $I_0$.  To isolate this UV divergence, we'd like to
isolate a term that (i) has the same UV divergence as $I_0$, (ii) is
analytically computable in dimensional regularization, and (iii) is
infrared convergent (since otherwise we'll just re-introduce an infrared
divergence when we isolate it).  Something which satisfies all these
requirements is the same integrals (\ref{eq:I0b}) of a three-dimensional theory
as above but with mass terms to cut off the infrared:
\begin {eqnarray}
   I_\Mir &\equiv&
      - {T^2\over2} \left(8\pi a\over m\right)^2
      \int_{\q\k} { 1 \over
         (\omega_q+\Mir) (\omega_k+\Mir) (\omega_{\q+\k}+\Mir)] }
\nonumber\\
   &=&
      - 4m^3T^2 \left(8\pi a\over m\right)^2
      \int_{\q\k} { 1 \over
         (q^2+\mir^2) (k^2+\mir^2) (|\q+\k|^2+\mir^2) }
   \,,
\label{eq:IMir}
\end {eqnarray}
where $\Mir = \mir^2/2m$ is an arbitrary frequency scale.
Our strategy will then be to rewrite our
current expression (\ref{eq:Pidn3}) as
\begin {equation}
   \Pi_\psi^{\rm(d)}(0) = \left[
      \left(8\pi a\over m\right)^2
      \int_{\k\l}
         {n(\omega_k)\,n(\omega_l) - n_0(\omega_k)\,n_0(\omega_l)
         \over \omega_l - \omega_{\k-\l} - \omega_k}
      + I_\Mir \right] - I_\Mir
      + {\cal O}(\eps) .
\label {eq:Pidn4}
\end {equation}
To put the first $I_\Mir$ term in a form similar to the integral shown
explicitly in (\ref{eq:Pidn4}), one may replace $\omega$ by $\omega+\Mir$
in our early discussion of $\Pi_\psi^{(d)}(0)$ and take the $\beta \to 0$
limit in all integrands to get the following analogy to
(\ref{eq:Pidn}):
\begin {eqnarray}
   I_\Mir &=&
      - {1\over2} \left(8\pi a\over m\right)^2
      \int_{\q\k\l}
         {[n_0(\omega_q+\Mir)\,n_0(\omega_k+\Mir)
            - 2 n_0(\omega_k+\Mir)\,n_0(\omega_l+\Mir)]
         \over (\omega_l+\Mir) - (\omega_q+\Mir) - (\omega_k+\Mir)}
\nonumber\\ && \hspace{15em} \times
         (2\pi)^d \delta^{(d)}(\l-\q-\k) + {\cal O}(\eps) .
\end {eqnarray}

We then obtain
\begin {eqnarray}
   \Pi_\psi^{\rm(d)}(0) &=&
      \left(8\pi a\over m\right)^2 \Biggl\{
      \int_{\k\l} \left[
         {n(\omega_k)\,n(\omega_l) - n_0(\omega_k)\,n_0(\omega_l)
              \over \omega_l - \omega_{\k-\l} - \omega_k}
         + {n_0(\omega_k+\Mir)\,n_0(\omega_l+\Mir)
              \over \omega_l - \omega_{\k-\l} - \omega_k - \Mir}
      \right]
\nonumber\\ && \hspace{5em}
      - {1\over2} \int_{\k\l} {n_0(\omega_{\k-\l}+\Mir)\,n_0(\omega_k+\Mir)
              \over \omega_l - \omega_{\k-\l} - \omega_k - \Mir}
     \Biggr\}
      - I_\Mir
      + {\cal O}(\eps) ,
\label {eq:Pidn5}
\end {eqnarray}
The first integral (with its implicit $\PP$ prescription) is now both
infrared and ultraviolet convergent and can now be evaluated in exactly
$d{=}3$ dimensions.  The second integral is convergent
as well.  So fix $d{=}3$ in these integrals, scale out the parameters,
and do the angular integrations using the $\PP$ prescription.  This puts
the integrals in a form appropriate for straightforward numerical evaluation.
The result for the dimensionally regulated integral (\ref{eq:IMir}) for
$I_\Mir$ is
\begin {equation}
      \int_{\q\k} { 1 \over
         (q^2+\mir^2) (k^2+\mir^2) (|\q+\k|^2+\mir^2) }
      = {1\over (4\pi)^2}\left[
            {1\over2\eps} + \ln {\bar\ren\over 3\mir} + {1\over2}
        \right]
        + {\cal O}(\eps) ,
\end {equation}
which can be extracted from the general $d$ result of Ref.\
\cite{ford} or the $\eps \to 0$ analysis in Ref.\ \cite{frks}.
Putting everything together,
\begin {equation}
   \Pi_\psi^{\rm(d)}(0) =
    {32\pi a^2 T \over\lambda^2}
        \left[{1\over2\eps} + \ln(\bar\ren\lambda) + C_1\right] ,
\end {equation}
where $C_1$ is the numerical constant
\begin {mathletters}%
\label {eq:C1}%
\begin {eqnarray}
   C_1 =
   {1-\ln(36\pi\bar\Mir)\over 2}
   &+& {2\over\pi^2} \int_0^\infty dk\>dl \Biggl[
        \left({kl\over (e^{k^2}-1)(e^{l^2}-1)} - {1\over k l}\right)
        \ln \left|k-l\over k+l\right|
\nonumber\\ && \hspace{8em}
        + {kl\over (k^2+\bar\Mir)(l^2+\bar\Mir)}
        \ln \left|\bar\Mir+2k(k-l)\over\bar\Mir+2k(k+l)\right|
     \Biggr]
\nonumber\\
   &+& {1\over\pi^2} \int_0^\infty dq\>dk\>
       {qk\over (q^2+\bar\Mir)(k^2+\bar\Mir)}
       \ln\left|\bar\Mir+2qk\over\bar\Mir-2qk\right|
   ,
\end {eqnarray}%
which is independent of the choice of the dimensionless number
$\bar\Mir \equiv \beta\Mir$.
Numerical evaluation of the integrals gives
\begin {equation}
   C_1 \simeq -0.54410 .
\end {equation}%
\end {mathletters}%
(Since completion of this work, a somewhat more compact formula for $C_1$
has been derived in Ref.\ \cite{boselog}.)

We should mention that it is possible, at a formal level, to turn the
original unregulated integral of (\ref{eq:Pidn2}) into a double sum,
similar to the sums appearing the earlier formula (\ref{eq:nresult})
for the density, by using methods similar
to those reviewed in Appendix \ref{app:P}.
However, the infrared divergence
of $\Pi_\psi^{\rm(d)}(0)$ would manifest as $i,j\to\infty$ divergences of
these sums.  We found it easier to handle the infrared issues in the integral
form than in the summation form.  This is the  only reason why our
treatment of $\Pi_\psi(0)$ superficially looks so dissimilar, in
final form, to our treatment of pressure and density in Section
\ref{sec:N}.



\subsubsection {Final result for $r$}

Combining our results for the pieces of $\Pi_\psi(0)$ with the matching formula
(\ref{eq:rmatch2}) for $r$, we obtain
\begin {equation}
   r_\bare = -2m \mu + 2mT\left\{ {4 a\over\lambda} \zeta(\threehalf)
       + {32\pi a^2\over\lambda^2} \left[
            {1\over2\eps} + \ln(\bar\ren\lambda) + C_1 \right] \right\}
       + {\cal O}(a^3) + {\cal O}(\mu-\muc) .
\end {equation}
Comparing to the expression (\ref{eq:rMSbar}) for the \MSbar\ definition
of $r$, and using the leading-order result (\ref{eq:u}) for $u$,
\begin {equation}
   r_\msbar(\bar\ren) =
     -2m \mu + 2mT\left\{ {4 a\over\lambda} \zeta(\threehalf)
       + {32\pi a^2\over\lambda^2} \left[
            \ln(\bar\ren\lambda) + C_1 \right] \right\}
       + {\cal O}(a^3) + {\cal O}(\mu-\muc) .
\label{eq:rfinal}
\end {equation}


\subsection {Final result for $\muc$}

We can now solve (\ref{eq:rfinal}) for the coefficients in the
expansion
\begin {mathletters}%
\label {eq:muc2nd}%
\begin {equation}
   \muc^{\phantom{(1)}} = \left[\muc^{(1)}{a\over\lambda}
             + \muc^{(2)}\left(a\over\lambda\right)^2 + \cdots \right] \kB T
\end {equation}
of $\muc$.  The first-order result, well known in the literature, is
\begin {equation}
   \muc^{(1)} = 4\,\zeta(\threehalf) ,
\end {equation}
and is a simple consequence of the vanishing susceptibility as depicted
in Fig.\ \ref{fig:cancel}.  The second-order coefficient is
\begin {equation}
   \muc^{(2)} = 32\pi \left[ \ln(\bar\ren\lambda) + C_1
                  - 72\pi^2 \, {r_{\c,\msbar}(\bar\ren,u) \over u^2} \right],
\end {equation}%
\end {mathletters}%
where $r_{\c,\msbar}(\bar\ren,u)$ is the critical value of $r_\msbar$
for a given choice of coupling $u$ and renormalization scale $\bar\ren$.
The only dimensionful scale of the three-dimensional theory at its
critical point is $u$, and so one should pick the renormalization
scale $\bar\ren$ of order $u$.
Note that the critical value $r_{c,\msbar}$ is then proportional to
$u^2$ by dimensional analysis.
Taking $\bar\ren = u/3$ for definiteness,
and because that was the choice made in presenting lattice simulation
results in ref.\ \cite{boselat2}, we have
\begin {equation}
   \muc^{(2)} =
     32\pi \left[ \ln\left(32\pi^2a\over\lambda\right) + C_1
     - 72 \pi^2 {\cal R} \right] ,
\label{eq:muc2ndB}
\end {equation}
where the dimensionless constant
\begin {mathletters}%
\label {eq:R}
\begin {equation}
   {\cal R} \equiv {r_{\c,\msbar}(\bar\ren{=}u/3,u) \over u^2}
\end {equation}%
is non-perturbative and must be extracted from simulations of the
three-dimensional effective theory (\ref{eq:O2}).
The simulation result is, from Eq.\ (1.5) of Ref.\ \cite{boselat2},
\begin {equation}
   {\cal R} = 0.001920(2) .
\end {equation}%
\end {mathletters}%


\section {Final Result for $\Tc$}
\label {sec:final}

We can now combine the second-order result (\ref{eq:muc2ndB}) for $\muc$
with our earlier expression (\ref{eq:Tcxmu2}) for $\Tc$ to obtain
\begin {mathletters}
\label {eq:finalTc}
\begin {equation}
   \Tc = T_0 \left[ 1 + c_1 \, {a\over\lambda_0}
             + \left(c_2' \ln {a\over\lambda_0} + c_2''\right)
                        \left(a\over\lambda_0\right)^2
             + O\!\left(a\over\lambda_0\right)^3 \right] ,
\end {equation}%
with
\begin {equation}
   c_1 \simeq -3.426~032 ,
\end {equation}
\begin {equation}
   c_2'  = - {32\pi\,\zeta(2)\over 3\,\zeta(3)} \,,
\end {equation}
\begin {equation}
   c_2'' = C_2 - {32\pi\,\zeta(2)\over 3\,\zeta(3)} \left[
           \ln(32\pi^2) + C_1 - 72\pi^2{\cal R} 
         \right]
         \simeq -155.0
\end {equation}
\end {mathletters}%
and with $T_0$, $\lambda_0$ the ideal gas quantities given at the
end of Section \ref{sec:N}.  The constants $c_1$, $C_2$, $C_1$, and
${\cal R}$ are given by Eqs.\ (\ref{eq:c1}), (\ref{eq:C2}),
(\ref{eq:C1}), and (\ref{eq:R}) respectively.  All have been computed
perturbatively, except for ${\cal R}$, which is the non-perturbative
information extracted from lattice simulations.


\section {Yet higher order corrections}
\label {sec:higher}

We have based our discussion on the 3+1 dimensional theory (\ref{eq:L1})
of $\psi$ and the effective 3-dimensional theory (\ref{eq:O2}) of the
zero-frequency Matsubara modes.
Both of these theories are approximate and have corrections which we
have ignored, claiming them to be higher order than the order of interest.
In this section, we will briefly discuss the nature of those corrections.

Let's begin with the original 3+1 dimensional theory (\ref{eq:L1}) of $\psi$.
Among other things, this theory ignores (a) the energy-dependence of the
low-energy atomic scattering cross-section, and (b) the effects of
3-body collisions.  Braaten, Hammer, and Hermans \cite{bhh} give a nice
discussion of how to systematize the corrections to the low-energy
3+1 dimensional theory, discussing interactions that are progressively
more and more irrelevant at low energies.  The most important such
corrections are to supplement the Lagrangian (\ref{eq:L1}) by the additional
interactions
\begin {equation}
   \delta{\cal L} =
   - {\pi\hbar^2 a^2 \reff \over 2 m} \, |\grad(\psi^*\psi)|^2
   - B (\psi^*\psi)^3 .
\label {eq:Lcorr}
\end {equation}
$B$ parameterizes the amplitude for 3-body\
collisions.
$\reff$ is the effective range of the 2-body
scattering problem and parameterizes the linear term in the
energy dependence of the cross-section at low energy.
The importance of the $\reff$ term grows with energy, which turns out
to mean that its leading effect on the critical temperature or the chemical
potential is not
infrared dominated and can be treated perturbatively.

The parametric size of the leading-order effects of these corrections can be
estimated in a very simple way by comparing them to the usual quartic
term $(\psi^*\psi)^2$.  At {\it leading}\/ order, the effects of
$(\psi^*\psi)^2$ on the quantities computed in this paper [$\mu_c$ for
a homogeneous gas and $N(T_\c)$ for a trapped gas] were dominated by
momentum scales of order $\bar k \sim 1/\lambda$
(as opposed to the infrared scale $u$).
Relative to the $a (\psi^*\psi)^2$ interaction, one would expect that the
leading-order effects of the $a^2 \reff |\grad(\psi^*\psi)|^2$ interaction of
(\ref{eq:Lcorr}) should therefore be
suppressed by $a \reff \bar k^2 \sim a \reff / \lambda^2$.
Near the transition, this is down by two powers of the typical inter-particle
separation $l \sim \bar n^{-1/3} \sim N^{-1/6} \aho$ discussed in the
introduction (since $\lambda \sim l$ at $\Tc$), whereas the second-order
effects computed in this paper are down only by one power,
compared to the leading-order effect of interactions.
Again relative to the $(\psi^*\psi)^2$ interaction, one would expect that
the 3-body $(\psi^*\psi)^3$ interaction of (\ref{eq:Lcorr}) is down by
a factor of
$(B m/\hbar^2 a) \psi^*\psi \sim (B m/\hbar^2 a) n$,
which is down by three powers of $l$ since
$n \sim l^{-3}$.  The moral is that corrections to the original 3+1
dimensional Lagrangian (\ref{eq:L1}) do not matter for a second-order
calculation of $\Tc$ for a dilute trapped gas, that the result at third
order would depend on the effective range $\reff$ and not just the
scattering length $a$, and that the result at fourth order would depend on
the 3-body scattering rate as well.

\begin {figure}
\vbox{
   \begin {center}
      \epsfig{file=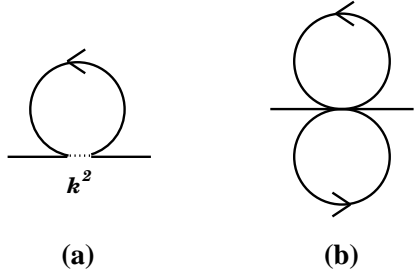,scale=.5}
   \end {center}
   \caption{
       Leading-order corrections to the $\Pi_\psi(0)$ (and hence the
       determination of $\muc$) due to (a) the effective range
       and (b) the 3-body scattering terms of (\ref{eq:Lcorr}).
       The dotted line represents the momentum flow $k$ in the
       $|\grad(\phi^*\psi)|^2$ vertex.
       \label {fig:mucorr}
   }
}
\end {figure}

\begin {figure}
\vbox{
   \begin {center}
      \epsfig{file=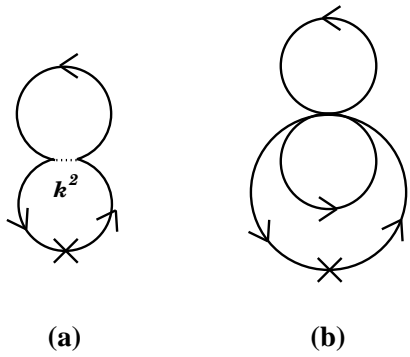,scale=.5}
   \end {center}
   \caption{
       As Fig.\ \ref{fig:mucorr} but showing corrections to the density
       $n(T,\mu)$.
       \label {fig:ncorr}
   }
}
\end {figure}

One can also verify the above analysis by a consideration of the
leading-order diagrams involving a given correction from (\ref{eq:Lcorr}).
Fig.\ \ref{fig:mucorr} shows diagrams contributing to $\muc$ and
Fig.\ \ref{fig:ncorr} those%
\footnote{
  The effect of $\reff$
  represented by Fig.\ \ref{fig:ncorr}a has been considered
  historically  in discussions of $\Delta\Tc$ for a homogeneous Bose gas
  \cite{luban}.
  Those discussions completely missed the dominant contributions
  to $\Delta\Tc$.  They also did not use the more general
  language of effective ranges
  but implicitly used Born approximation
  to express $\reff$ in terms of the 2-body potential.
}
contributing to $n(T,\mu)$.  As an example,
the diagram of Fig.\ \ref{fig:mucorr}a gives a contribution to the
chemical potential proportional to
\begin {equation}
   \delta\mu \sim {a^2\reff\over\hbar m} \sumint {k^2\over i k_0 + \omega_l} .
\end {equation}
The diagram is not dominated by infrared momenta, and so the
perturbative treatment is justified.
The dominant wave numbers are $\bar k \sim 1/\lambda$, as claimed above,
corresponding to energies $\kB T$ and frequencies
$\omega_{\bar k} \sim \kB T/\hbar$.
The result is that
\begin {equation}
   \delta\mu \sim {a^2\reff\over\hbar m} \,
                  {\kB T \bar k^5 \over \omega_{\bar k}}
              \sim {a^2\reff\over\lambda^3} \, \kB T ,
\label {eq:dmucorr}
\end {equation}
where the $\bar k^5$ comes from the $k^2 \> d^3k$ in the integral.
Compared to the leading-order result $O(\kB T a/\lambda)$ for
the chemical potential, (\ref{eq:dmucorr}) is down by
$O(a \reff/\lambda^2)$, just as we argued more simply above.

\begin {figure}
\vbox{
   \begin {center}
      \epsfig{file=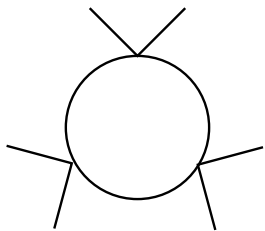,scale=.5}
   \end {center}
   \caption{
       An effective $(\phi^*\phi)^3$ interaction of zero modes generated by
       a loop of non-zero modes.
       \label {fig:phi6}
   }
}
\end {figure}

Finally, even ignoring corrections to the original 3+1 dimensional
theory, there will still be corrections to the effective 3 dimensional
theory (\ref{eq:O2}) of the zero modes.  One might worry in particular
about a $(\phi^*\phi)^3$
interaction between the zero modes, which is a marginal
interaction in three dimensions.  Such an effective
interaction can be induced by
diagrams such as Fig.\ \ref{fig:phi6} in the 3+1 dimensional
theory, where the external lines
are zero-modes and the internal lines are non-zero modes.
However, the non-zero modes are infrared insensitive and are
dominated by frequencies of order $\kB T/\hbar$ and momenta of order
$\bar k \sim 1/\lambda$.  Power counting Fig.\ \ref{fig:phi6} then gives
an interaction in the effective 3 dimensional theory of order
\begin {equation}
   \delta {\cal L}_\phi \sim u^3 \lambda^3 (\phi^*\phi)^3 ,
\label{eq:O2corr}
\end {equation}
where the $u^3$ can be understood as arising from the three vertices in
Fig.\ \ref{fig:phi6} and then the $\lambda^3$ from dimensional analysis based
on the dominant momentum scale.

Now consider the effect of the vertex (\ref{eq:O2corr}) on the infrared
physics at momentum scales $p \sim 1/u$, to which the 3-dimensional
effective theory is intended to be applied.  At that scale, the
$(\phi^*\phi)^2$ interaction can no longer be treated perturbatively and, by
dimensional analysis, the fluctuations in $\phi$ are of order $u^{1/2}$.
The relative importance of the $(\phi^*\phi)^3$ term at the infrared scale
$p \sim 1/u$ is then
\begin {equation}
   {u^3 \lambda^3 (\phi^*\phi)^3 \over u (\phi^*\phi)^2}
   \sim u^3 \lambda^3
   \sim {a^3 \over \lambda^3} .
\end {equation}
The contributions of the effective
$(\phi^*\phi)^3$ operator is therefore down by
three powers of $l^{-1} \sim \lambda^{-1}$ compared to those contributions
we have included in this paper.
Other corrections to the three-dimensional theory are similarly suppressed.

\section {How wide is a wide trap?}
\label {sec:wide}

We have assumed throughout that the trap is arbitrarily wide
($\omega_x,\omega_y,\omega_z \to 0$ with $N\omega_x\omega_y\omega_z$ fixed).
We will now take a moment to explain parametrically how wide ``wide enough''
is for our second-order results to be valid.
Our second-order result for $\muc$ depends on non-perturbative physics
near the center of the trap, and we treated the trap as flat over the
wavelength $1/u$ of such physics.
The trap must therefore be wide enough that this wavelength fits
comfortably inside the region of the trapped gas that is nearly
critical, whose size we labeled $\Rnp$ in Section \ref{sec:gasintrap}.

First consider the case $\omega_x \sim \omega_y \sim \omega_z$.
Using Table \ref{tab:scales}, the condition $1/u \ll \Rnp$ can be
translated into $l \ll N^{1/6} a$.  Combining this with the basic
diluteness assumption $a \ll l$ of our analysis, we then require
\begin {equation}
   a \ll l \ll N^{1/6} a .
\end {equation}
This shows only the parametric dependence, and we have made no attempt to
estimate numerical factors.

For a very anisotropic trap, the strongest constraint will come from
requiring the narrowest direction of the near-critical region to be
larger than $1/u$.  Let $\omegamax$ be the largest of $\omega_x$,
$\omega_y$, and $\omega_z$.  Repeating the analysis of Section
\ref{sec:gasintrap} then gives the corresponding value of $\Rnp$ in that
direction as $\hbar a/m\omegamax \lambda^2 \sim N^{1/3} a \omegaho/\omegamax$.
So the condition is
\begin {equation}
  a \ll l \ll N^{1/6} a \, \sqrt{\omegaho\over\omegamax} \, .
\label {eq:condition}
\end {equation}
These constraints may be translated into other variables using
$l \sim \bar n^{-1/2} \sim \lambda \sim N^{-1/6} \aho$.

\begin {figure}
\vbox{
   \begin {center}
      \epsfig{file=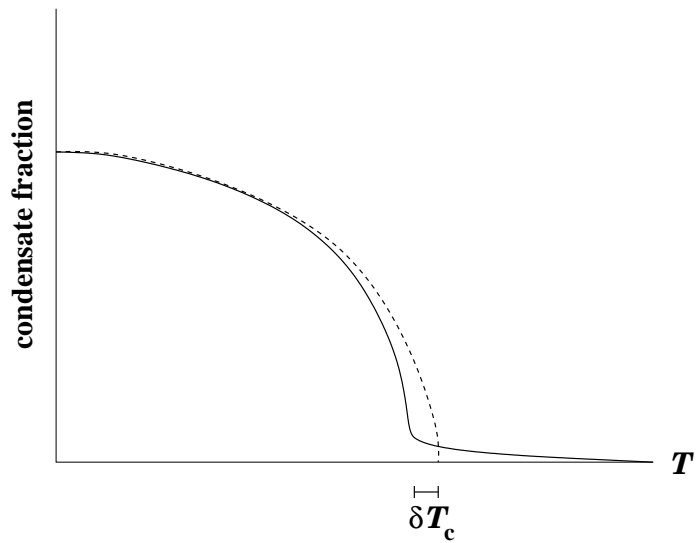,scale=.5}
   \end {center}
   \caption{
       A schematic depiction of finite-size effects on the BEC phase
       transition in the ideal-gas limit.
       The dashed curve indicates the infinite-volume transition.
       \label {fig:finiteN}
   }
}
\end {figure}

This condition on the size of the trap can also be summarized as a comparison
of the uncertainty in the value of $T_\c$ due to finite size effects vs.\
the resolution with which we have computed $T_\c$ in our second-order formula
(\ref{eq:finalTc}).  Finite size effects round off the non-analyticity of
the infinite-size transition, as depicted in Fig.\ \ref{fig:finiteN}.
A standard result from the literature is that, below the rounded transition,
finite-size effects create the appearance of a transition
temperature shifted by \cite{review,finitesize}
\begin {equation}
   {\delta\Tc\over T_0} \simeq
     - {\zeta(2)\over 2\, \zeta(3)^{2/3}} \, {\bar\omega\over\omegaho}
         \, N^{-1/3}
\label {eq:finiteN}
\end {equation}
from the infinite-volume value (in the ideal gas approximation), as
depicted in the figure.
Here, $\bar\omega$ is the arithmetic mean
\begin {equation}
   \bar\omega \equiv {\omega_x+\omega_y+\omega_z \over 3} \, .
\end {equation}
The condition that this finite-size
effect on the transition be small compared to the relative $O(a^2/\lambda_0^2)$
correction to $T_0$ that we have presented in (\ref{eq:finalTc}) happens to
be the same, parametrically, as the right-hand condition in
(\ref{eq:condition}).

\section {Conclusion}
\label{sec:conclusion}

   The relative size of the second-order effect in our final result
(\ref{eq:finalTc}) for $\Tc$ obviously depends on the diluteness of the
gas and the value of the scattering length, which will vary from experiment
to experiment.  However, just for fun, let us put numbers to the size of
the effect for one specific experimental study of $\Tc$ that has
appeared in the literature.
The 1996 experiment of Ensher {\it et al.}\ \cite{ensher} found
$\Delta\Tc/T_0 = -0.06 \pm 0.05$ for dilute gases of roughly
$N=40,000$ ${}^{87}$Rb atoms in the $F=2$ hyperfine state,
trapped with
$\nu_z = 373$ Hz, $\omega_z = 2\pi\nu_z$, and
$\omega_x=\omega_y=\omega_z/\sqrt{8}$.
The relevant scattering length is $a = (103 \pm 5) \, a_0$
\cite {julienne},
where $a_0 = 0.0529177$ nm is the Bohr radius.
(See also Ref.\ \cite{boesten}.)
These parameters correspond to $a/\lambda_0 \simeq 0.016$.
For an arbitrarily wide trap, this would translate into a first-order
correction to $\Tc$ of roughly $-5.4$\% and a second-order correction
of roughly $+0.9$\%.  For the actual trap, however, the corrections
(\ref{eq:finiteN}) due to finite-size effects are roughly $-2.4$\%.
The fact that this is larger in magnitude than the second-order correction
leads us to suspect that this particular trap may not be wide enough for the
second-order result to be trusted, as was discussed in Section
\ref{sec:wide}.


\section* {ACKNOWLEDGMENTS}

We thank Eric Braaten, Eric Cornell, Cass Sackett,
and Guy Moore for useful discussions.
This work was supported, in part, by the U.S. Department
of Energy under Grant No.\ DE-FG02-97ER41027.

\appendix


\section {Field theory rederivation of
         \boldmath$P(T,\mu)$}
\label {app:P}

\begin {figure}
\vbox{
   \begin {center}
      \epsfig{file=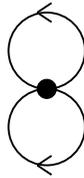, scale=.4}
   \end {center}
   \caption{
       An additional perturbative diagram contributing to the pressure
       at second order.  The fat black dot represents the one-loop
       $(\psi^*\psi)^2$ renormalization counter-term, which vanishes
       in dimensional regularization.
       \label{fig:Pcounter}
   }
}
\end {figure}

The two diagrams which contribute to the pressure at second order
in perturbation
theory were shown in Fig.\ \ref{fig:P}(b,c).
Technically, there is also a third diagram,
Fig.\ \ref{fig:Pcounter}, which involves the one-loop
$(\psi^*\psi)^2$ counter-term (represented by the fat dot)
for renormalizing the linear UV divergence
of zero-temperature, zero-density $2 \to 2$ scattering at second order.
However, we shall use dimensional regularization, for which this
counter-term vanishes, as discussed in Section \ref{sec:UV}.
The Feynman rules are given in Table \ref{tab:feyn2}.
They are the same as in Table \ref{tab:feyn} except that we are not
treating the chemical potential $\mu$ as a perturbation in this
context.  We work in units where $\hbar=\kB=1$.

\begin{table}[t]
\begin {center}
\tabcolsep=8pt

\begin {tabular}{|c|c|}             \hline
  \centering{$\vcenter{\hbox{\epsfig{file=figFprop.eps,scale=.3}}}$}
  & $\displaystyle{1\over i k_0 + \omega_\k - \mu}$
\\
  \centering{$\vcenter{\hbox{\epsfig{file=figFvertex.eps,scale=.3}}}$}
  & $\displaystyle{-{8\pi a\over m}}$
\\ \hline
\end {tabular}
\end {center}
\caption
    {%
    \label {tab:feyn2}
       Feynman rules appropriate for standard perturbation theory in
       the 3+1 dimensional theory.
    }%
\end{table}


\subsection {The basketball diagram}

Let's start with diagram (c) of Fig.\ \ref{fig:P}.
The corresponding
contribution to the pressure $P = (\beta V)^{-1}\ln Z$ is%
\footnote{
   $V P_{\rm c}$ corresponds to $\Q'/\Q^{(0)}$ of Ref.\ \cite{n1}.
   The contribution $V P_{\rm b}$ from diagram b corresponds to
   $[\Q''/\Q^{(0)}] - {1\over2} [\Q^{(1)}/\Q^{(0)}]^2$.
}
\begin {equation}
   P_{\rm c} = 
    {1\over8} \left(-8\pi a\over m\right)^2
    \sumint_{PQKL}
    {\beta \, \delta_{p_0+q_0-k_0-l_0} \,(2\pi)^3\delta^{(3)}(\p+\q-\k-\l)
    \over (ip_0+\tomega_p)
          (iq_0+\tomega_q)
          (ik_0+\tomega_k)
          (il_0+\tomega_l)}
   \,,
\end {equation}
where we have introduced the shorthand notation
\begin {equation}
   \tomega_k \equiv \omega_k - \mu = {k^2\over 2m} - \mu ,
\end {equation}
and where the summation-integration sign is defined in
(\ref{eq:defsumint}).
We now use standard tricks to evaluate the frequency sums.%
\footnote{
   See, for example, Section 5.5.1 of Ref.\ \cite{kapusta}.
}
Specifically, rewrite the frequency Kronecker $\delta$ as an
integral of exponentials, and factorize the expression into independent
sums:
\begin {equation}
   P_{\rm c} = {8 \pi^2 a^2\over m^2} \int_0^\beta d\alpha \>
        \sumint_P {e^{-i\alpha p_0} \over i p_0 + \tomega_p}
        \sumint_Q {e^{-i\alpha q_0} \over i q_0 + \tomega_q}
        \sumint_K {e^{+i\alpha k_0} \over i k_0 + \tomega_k}
        \sumint_L {e^{+i\alpha l_0} \over i l_0 + \tomega_l}
        \, (2\pi)^3 \delta^{(3)}(\p+\q-\k-\l)
\end {equation}
Then we use the standard frequency sums
\begin {mathletters}
\label {eq:fsums}
\begin {eqnarray}
   T \sum_{p_0} {e^{-i\alpha p_0} \over i p_0 + \omega}
       &=& n(\omega) \, e^{\alpha\omega} ,
\\
   T \sum_{p_0} {e^{+i\alpha p_0} \over i p_0 + \omega}
       &=& n(\omega) \, e^{(\beta-\alpha)\omega} ,
\end {eqnarray}
\end {mathletters}%
for $0<\alpha<\beta$.  The $\alpha$ integration is then trivial,
yielding
\begin {equation}
   P_{\rm c} = {8 \pi^2 a^2\over m^2} \int_{\p\q\k\l}
      n(\tomega_p)\,n(\tomega_q)\,n(\tomega_k)\,n(\tomega_l)\,
      {e^{\beta(\tomega_p+\tomega_q)} - e^{\beta(\tomega_k+\tomega_l)}
       \over \tomega_p+\tomega_q-\tomega_k-\tomega_l}
      \, (2\pi)^3 \delta^{(3)}(\p+\q-\k-\l)
   \,.
\label {eq:Pc1}
\end {equation}
Note that the zero of the denominator at
$\tomega_p+\tomega_q = \tomega_k+\tomega_l$ is canceled by a corresponding
zero of the numerator.  However, it will be useful to split the integral
into pieces that individually lack this cancellation, and so it is useful to
first introduce a redundant principal part ($\PP$) prescription in
(\ref{eq:Pc1}).
Making use of the identity
\begin {equation}
   n(\omega)\, e^{\beta\omega} = n(\omega) + 1 ,
\label{eq:nexp}
\end {equation}
expanding terms, and permuting integration variables, we can rewrite
(\ref{eq:Pc1}) as
\begin {eqnarray}
   P_{\rm c} &=& {32 \pi^2 a^2\over m^2} \int_{\p\q\k\l}
      \PP
      {
         n(\tomega_q)\,n(\tomega_k)\,n(\tomega_l)
         - \half n(\tomega_p)\,n(\tomega_q)
      \over
         \tomega_p+\tomega_q-\tomega_k-\tomega_l
      }
      \, (2\pi)^3 \delta^{(3)}(\p+\q-\k-\l)
   \,.
\nonumber\\
   &=& {32 \pi^2 a^2\over m^2} \int_{\p\q\k\l}
      \PP
      {
         n(\tomega_q)\,n(\tomega_k)\,n(\tomega_l)
         - \half n(\tomega_p)\,n(\tomega_q)
      \over
         \omega_p+\omega_q-\omega_k-\omega_l
      }
      \, (2\pi)^3 \delta^{(3)}(\p+\q-\k-\l)
   \,.
\label {eq:Pc2}
\end {eqnarray}

The term
\begin {equation}
  P_{\rm c,2}
   = {32 \pi^2 a^2\over m^2} \int_{\p\q\k\l}
      \PP
      {
         - \half n(\tomega_p)\,n(\tomega_q)
      \over
         \omega_p+\omega_q-\omega_k-\omega_l
      }
      \, (2\pi)^3 \delta^{(3)}(\p+\q-\k-\l)
   \,,
\label {eq:Pnn}
\end {equation}
involving just two $n$'s, has a linear UV divergence associated
with $k,l \to \infty$ with $p$ and $q$ fixed.  This is the divergence
that is canceled by the counter-term diagram of Fig.\ \ref{fig:Pcounter} for
generic regularization schemes and which dimensional regularization will
simply ignore.
In fact, the entire term $P_{\rm c,2}$ simply vanishes in dimensional
regularization, which can be seen by doing the $\k$ and $\l$ integrations
explicitly in $d$ spatial dimensions.  Defining $\s = \k - \half(\p+\q)$,
\begin {eqnarray}
   \int_{\k\l} \PP {(2\pi)^d \delta^{(d)}(\p+\q-\k-\l)
        \over \omega_p+\omega_q-\omega_k-\omega_{l}}
   &=& - m \int {d^ds\over(2\pi)^d} \> \PP {1\over s^2 -
        {1\over4}|\p-\q|^2}.
\label {eq:nn1}
\end {eqnarray}
It's convenient to re-express the principal part in terms of
infinitesimals,
using (\ref{eq:PP}), before doing the $\s$ integration.
The integral ({\ref{eq:nn1}) then yields
\begin {eqnarray}
&&
   - { m \Gamma\left(1-{d\over 2}\right) \over (4\pi)^{d/2} } \,
         \Re \left[ - \fourth|\p-\q|^2 + i 0^\pm \right]^{(d-2)/2}
\nonumber\\ && \qquad
   =
   - { m \Gamma\left(1-{d\over 2}\right) \over (4\pi)^{d/2} }
         \left[ \fourth|\p-\q|^2 \right]^{(d-2)/2}
         \cos\left((d-2)\pi\over 2\right) .
\end {eqnarray}
Analytic continuation to $d=3$ yields zero:
\begin {eqnarray}
   \int_{\k\l} \PP {(2\pi)^d \delta^{(d)}(\p+\q-\k-\l)
        \over \omega_p+\omega_q-\omega_k-\omega_{l}}
   = {\cal O}(\eps) ,
\label{eq:zero}
\end {eqnarray}
where $\eps = 3-d$.
To conclude that the contribution (\ref{eq:Pnn}) to the pressure vanishes
in dimensional regularization,
one must also check that the final $\p$ and $\q$ integrals with the
${\cal O}(\eps)$ integrand do not diverge, since divergences could possibly
generate a $1/\eps$ singularity to cancel the $O(\eps)$ behavior of the
integrand.  However, the UV is cut off by the distribution functions
$n(\omega_p)$ and $n(\omega_q)$ in (\ref{eq:Pnn}), and so this is not
an issue.

We are left with only the term of (\ref{eq:Pc2}) with three $n$'s:
\begin {equation}
   P_{\rm c} = {32\pi^2 a^2\over m^2} \int_{\p\q\k\l}
      \PP
      {
         n(\tomega_q)\,n(\tomega_k)\,n(\tomega_l)
      \over
         \omega_p+\omega_q-\omega_k-\omega_l
      }
      \, (2\pi)^3 \delta^{(3)}(\p+\q-\k-\l)
   \,.
\end {equation}
This reproduces eq.\ (A15) of Huang, Yang, and Luttinger \cite{n1}.
Since their subsequent discussion of evaluating this integral is somewhat
telegraphic, we will present our own method.
First, expand the distribution functions $n(\tomega)=n(\omega-\mu)$ in
powers of fugacity $z = \exp(\beta\mu)$:
\begin {equation}
   P_{\rm c} = {32 \pi^2 a^2\over m^2}
      \sum_{a=1}^\infty \sum_{b=1}^\infty \sum_{c=1}^\infty
      z^{a+b+c}
      \int_{\p\q\k\l}
      \PP
      {
         e^{-a\beta q^2/2m} e^{-b\beta k^2/2m} e^{-c\beta l^2/2m}
      \over
         \omega_p+\omega_q-\omega_k-\omega_l
      }
      \, (2\pi)^3 \delta^{(3)}(\p+\q-\k-\l)
   \,.
\end {equation}
Rescaling all momenta by $\sqrt{\beta/m}$ to make them dimensionless gives
\begin {equation}
   P_{\rm c} = {8 a^2 T\over\lambda^5}
      \sum_{a=1}^\infty \sum_{b=1}^\infty \sum_{c=1}^\infty
      z^{a+b+c} I_{abc} ,
\label {eq:Pcsemifinal1}
\end {equation}
\begin {equation}
   I_{abc} \equiv (2\pi)^{9/2}
      \int_{\p\q\k\l}
      \PP
      {
         e^{-a q^2/2} e^{-b k^2/2} e^{-c l^2/2}
      \over
         {1\over2}(p^2+q^2-k^2-l^2)
      }
      \, (2\pi)^3 \delta^{(3)}(\p+\q-\k-\l)
   \,.
\end {equation}

For the sake of justifying later manipulations, it is convenient to
introduce a redundant $\exp(-0^+ p^2)$ convergence factor into
the integral defining $I_{abc}$.
We will evaluate $I_{abc}$ by representing the energy denominator and the
$\delta$ function as integrals of exponentials.
Using the infinitesimal version (\ref{eq:PP}) of the principal part
prescription, we write
\begin {eqnarray}
   I_{abc} &=& (2\pi)^{9/2} \Re
      \int_{\p\q\k\l}
         e^{-0^+ p^2} e^{-a q^2/2} e^{-b k^2/2} e^{-c l^2/2}
\nonumber\\ && \qquad \times
      \int_0^{i\infty} d\lambda\>
         e^{-(p^2+q^2-k^2-l^2-i0^+)\lambda/2}
      \int d^3x\>
         e^{i\x\cdot(\p+\q-\k-\l)} .
\end {eqnarray}
The $\p$, $\q$, $\k$, and $\l$ integrations are now simple Gaussian integrals,
yielding
\begin {eqnarray}
   I_{abc} &=& (2\pi)^{-3/2} \Re
      \int_0^{i\infty} d\lambda \> e^{i 0^+\lambda}
      \int d^3x\>
      (0^++\lambda)^{-3/2} (a+\lambda)^{-3/2}
      (b-\lambda)^{-3/2} (c-\lambda)^{-3/2}
\nonumber\\ && \qquad \times
      \exp\left[-{x^2\over2}\left(
         {1\over 0^++\lambda} + {1\over a+\lambda}
         + {1\over b-\lambda} + {1\over c-\lambda}
      \right) \right] .
\end {eqnarray}
The $\exp(i0^+\lambda)$ prescription is now redundant and can be dropped.
It's also convenient to change integration variables from $\lambda$
to $\lambda+0^+$ in order to remove the remaining $0^+$ prescription from
the integrand (noting that $a,b,c \not= 0$).
The $\x$ integral is Gaussian and yields
\begin {equation}
   I_{abc} = \Re
      \int_{0^+}^{i\infty+0^+} d\lambda
      \left[ a b c + 2 b c \lambda - (a+b+c) \lambda^2 \right]^{-3/2} .
\end {equation}
The final  integral is straightforward and gives
\begin {equation}
   I_{abc} = {1\over (a+b)(a+c) (abc)^{1/2}} \,.
\label {eq:Pcfinal2}
\end {equation}
The final result for this contribution to the pressure is
\begin {equation}
   P_{\rm c} = {8 a^2 T\over\lambda^5}
      \sum_{a=1}^\infty \sum_{b=1}^\infty \sum_{c=1}^\infty
      {z^{a+b+c} \over (a+b)(a+c) (abc)^{1/2}} .
\label {eq:Pcfinal }
\end {equation}

\subsection {The three-circle diagram}

The other second-order diagram, Fig.\ \ref{fig:P}b, is trivial in
comparison.  It's contribution to the pressure is
\begin {equation}
   P_{\rm b} =
     {1\over2} \left(-8\pi a\over m\right)^2
       \sumint_P {1\over (i p_0 + \tomega_p)^2}
       \left( \sumint{1\over i q_0 + \tomega_q}\right)^2 .
\end {equation}
One of the summation-integrals is
\begin {equation}
   \sumint_Q {1\over iq_0 + \tomega_q}
   = \int_\q \left[n(\tilde\omega_q)+\half\right]
   = \int_\q n(\tomega_q)
   = \left(m \over 2\pi\beta \right)^{3/2} \Li_{3/2}(z) .
\end {equation}
The other is easily obtained by differentiating with respect to $\mu$:
\begin {equation}
   \sumint_P {1\over (ip_0 + \tomega_p)^2}
   = \beta \left(m \over 2\pi\beta \right)^{3/2} \Li_{1/2}(z) .
\end {equation}
So,
\begin {equation}
   P_{\rm b} = {8 a^2 T \over \lambda^5} [\Li_{3/2}(z)]^2 \Li_{1/2}(z) .
\end {equation}

Putting $P_{\rm b}$ and $P_{\rm c}$ together
gives the total second-order contribution to the pressure which
appears in (\ref{eq:P}).  The first-order contribution of Fig.\ \ref{fig:P}a
is easily evaluated in a similar manner.


\section {Small $\bar\mu$ expansion of $N$}
\label {app:Nx}

\subsection {The expansion}

Consider the $(a/\lambda)^2$ term in the expansion
(\ref{eq:Nx}) for $N$.  First consider the term proportional to
\begin {equation}
   \sum_{ijk} {\bar z^{i+j+k}\over (ij)^{3/2} k^{1/2} (i+j+k)^{1/2}} \,.
\end {equation}
Because of the explicit factor of $a^2$, one might naively
think one could use the order $a^0$ result $\bar z \simeq 1$ for $\bar z$.
But this would give
\begin {equation}
   \sum_{ijk} {1\over (ij)^{3/2} k^{1/2} (i+j+k)^{1/2}} \,,
\label{eq:sum1}
\end {equation}
which has logarithmic divergences associated with $k \to \infty$ with
$i$ and $j$ fixed.  We can isolate these divergences by rewriting the original
sum as
\begin {equation}
   \sum_{ijk} {\bar z^{i+j+k}\over (ij)^{3/2} k^{1/2} (i+j+k)^{1/2}}
   = \sum_{ijk} \left[
         {\bar z^{i+j+k}\over (ij)^{3/2} k^{1/2} (i+j+k)^{1/2}}
         - {\bar z^k\over (ij)^{3/2} k } \right]
      + \sum_{ijk} {\bar z^k\over (ij)^{3/2} k }
   \,.
\end {equation}
We can now safely set $\bar z$ to 1 in the first sum on the left-hand side.
The second sum is easy, giving $-\zeta(\threehalf)^2 \ln(1-\bar z)$.
The small $\bar\mu$ result is then
\begin {eqnarray}
&&
   \sum_{ijk} {\bar z^{i+j+k}\over (ij)^{3/2} k^{1/2} (i+j+k)^{1/2}}
\nonumber\\ && \qquad
   = \sum_{ijk} \left[
         {1\over (ij)^{3/2} k^{1/2} (i+j+k)^{1/2}}
         - {1\over (ij)^{3/2} k } \right]
      - \zeta(\threehalf)^2 \ln(-\beta\bar\mu)
      + O(\bar\mu)
   \,.
\label{eq:sum1x}
\end {eqnarray}

The sum associated with the $a/\lambda$ term of (\ref{eq:Nx}) must be
expanded to first order in $\bar\mu$, where it suffers a similar problem.
Naively,
\begin {equation}
   \sum_{ij} {\bar z^{i+j}\over (ij)^{3/2} (i+j)^{1/2}}
   = \sum_{ij} {1\over (ij)^{3/2} (i+j)^{1/2}}
     - \beta\bar\mu \sum_{ij} {(i+j)^{1/2} \over (ij)^{3/2}}
     + \cdots .
\end {equation}
The second term has logarithmic divergences associated with
(i) $i \to \infty$ with $j$ fixed, and, symmetrically, (ii) $j \to \infty$
with $i$ fixed.  Proceeding as before, we can isolate the divergent behavior
by writing
\begin {eqnarray}
&&
   \sum_{ij} {\bar z^{i+j}\over (ij)^{3/2} (i+j)^{1/2}}
\nonumber\\ && \qquad
   = \sum_{ij} \left[
           {\bar z^{i+j}\over (ij)^{3/2} (i+j)^{1/2}}
           - {\bar z^i \over i^2 j^{3/2}}
           - {\bar z^j \over i^{3/2} j^2} \right]
     + \sum_{ij} \left[
           {\bar z^i \over i^2 j^{3/2}}
           + {\bar z^j \over i^{3/2} j^2} \right] .
\end {eqnarray}
In the first sum, we can now safely replace $\bar z$ by
$1 + \beta\bar\mu + O(\bar\mu^2)$, and the second sum gives
\begin {equation}
   2 \zeta(\threehalf) \Li_2(\bar z)
   = 2 \zeta(\threehalf) \Bigl\{
           \zeta(2) + \left[-\ln(-\beta\bar\mu) + 1\right] \beta\bar\mu
     \Bigr\}
     + O(\bar\mu^2) .
\end {equation}
The final result for the expansion is then
\begin {eqnarray}
   \sum_{ij} {\bar z^{i+j}\over (ij)^{3/2} (i+j)^{1/2}}
   &=& \sum_{ij} {1\over (ij)^{3/2} (i+j)^{1/2}}
     + 2 \zeta(\threehalf) \, \beta\bar\mu [-\ln(-\beta\bar\mu)+1]
\nonumber\\ && \hspace{8em}
     + \beta\bar\mu \sum_{ij}
           {(i+j)^{1/2} - i^{1/2} - j^{1/2}\over (ij)^{3/2}}
     + O(\bar\mu^2)
   \,.
\nonumber\\
\label {eq:sum2x}
\end {eqnarray}

The last thing we need is the expansion of the sum in the
order $a^0$ term of (\ref{eq:N}), which is just
\begin {equation}
   \Li_3(\bar z) = \zeta(3) + \zeta(2) \beta\bar\mu
      + \half \left[ - \ln(-\beta\bar\mu) + \threehalf \right]
             (\beta \bar\mu)^2 
      + O(\bar\mu^3) .
\label{eq:sum3x}
\end {equation}
Combining the expansions (\ref{eq:sum1x}), (\ref{eq:sum2x}) and
(\ref{eq:sum3x}) of the sums with the expansion (\ref{eq:Nx}) of
$N$, we obtain
\begin {eqnarray}
   N &=& \left(\kB T \over \hbar\omegaho\right)^3 \Biggl\{
     \zeta(3)
     + \Biggl[
         \zeta(2) \, \beta\bar\mu
         - {2a\over\lambda} \sum_{ij} {1\over i^{3/2} j^{3/2} (i+j)^{1/2}}
       \Biggr]
\nonumber\\ &&
     + \Biggl[
         {\textstyle{3\over4}} (\beta\bar\mu)^2
         - {2a\over\lambda} \, \beta\bar\mu
              \sum_{ij} {(i+j)^{1/2}-i^{1/2}-j^{1/2}\over i^{3/2}j^{3/2}}
         - {4a\over\lambda}\,\beta\bar\mu\,\zeta(\threehalf)
\nonumber\\ && \qquad
         + 8 \left(a\over\lambda\right)^2
           \sum_{ijk} {1\over (ij)^{3/2} k^{1/2}} \left(
              {1 \over (i+j+k)^{1/2}} + {ij\over(i+k)(j+k)(i+j+k)^{1/2}}
              - {1\over k^{1/2}}
           \right)
       \Biggr]
\nonumber\\ &&
     - \half\left(\beta\bar\mu - {4 a \over \lambda}\zeta(\threehalf)\right)^2
       \ln(-\beta\bar\mu)
     + O\!\left({a\over\lambda},\beta\bar\mu\right)^3
   \Biggr\} .
\end {eqnarray}
If we now use the expansion ({\ref{eq:mucx}) of $\muc$, we obtain the
result (\ref{eq:Nx2}) presented in the main text.


\subsection {Cancellation of logarithms}

To understand the origins of the logarithms in the preceding analysis,
consider a straight, naive, perturbative expansion in $\bar\mu$.
Treating the $-\bar\mu \psi^* \psi$ term of the Lagrangian as a perturbation,
the logarithms then arise from the diagrams of Fig.\ \ref{fig:logs}.
Each diagram should be understood as evaluated at fixed $\x$, with
effective chemical potential $\mu = \bar\mu - V(\x)$, and then the
result of the diagram integrated over $\x$.  $\bar\mu$ is treated
perturbatively, while $V(\x)$ is not.  The imaginary-time
propagators in this perturbation
theory, derived from the action (\ref{eq:SI}), are
\begin {equation}
   G_0(p_0,\p) = {1 \over i p_0 + {p^2\over 2m} + V(\x)} ,
\end {equation}
where we now set $\hbar = 1$ for convenience.
The logarithms are produced by
the infrared behavior, near the center of the trap, of the loops drawn
large in the figure.  Specifically, it is the
$p_0{=}0$, $\p{\to}0$, $\x{\to}0$ behavior of these diagrams, where $p$ is
the loop momentum of those loops.
The small loops are UV dominated and so, to this order in the expansion
in $a$, are insensitive to $\x$ near the center of the trap.
The infrared divergences due to the large loops then produce the same
common factor for all diagrams:
\begin {equation}
   \int d^3 x \int d^3 p \> [G(0,\p)]^3
   \propto \int {d^6 q \over q^6} = \mbox{log divergent} ,
\end {equation}
where we've introduced the 6-dimensional phase-space vector
\begin {equation}
   \q = \left(
        {p_x\over\sqrt{2m}},
        {p_y\over\sqrt{2m}},
        {p_z\over\sqrt{2m}},
        {\sqrt{m\over2} \,\omega_x x},
        {\sqrt{m\over2} \,\omega_y y},
        {\sqrt{m\over2} \,\omega_z z} \right)
\end {equation}

\begin {figure}
\vbox{
   \begin {center}
      \epsfig{file=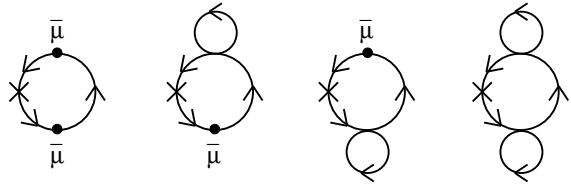,scale=.4}
   \end {center}
   \caption{
       Diagrams producing the infrared logarithm in the small $\bar\mu$
       expansion of $N$.  Each diagram should be understood as being
       evaluated with an effective chemical potential $\mu = \bar\mu - V(\x)$
       with $\bar\mu$ treated perturbatively.
       The dots represent the 2-point vertex coming from treating the
       $-\bar\mu\psi^*\psi$ term in the Lagrangian perturbatively.
       The crosses are as in Fig.\ \ref{fig:n}.
       \label {fig:logs}
   }
}
\end {figure}

The cancellation of these logarithms at the phase transition occurs
because, at the phase transition, the inverse susceptibility vanishes
at the center of the trap.
This condition is shown diagrammatically in Fig.\ \ref{fig:cancel},
which implies that the
logarithms generated by the diagrams of Fig.\ \ref{fig:logs} cancel
each other at the order of $a$ under consideration.


\section {Numerical results for sums}
\label {app:sums}

The following sums were computed numerically using iterative application of
the Euler-MacLaurin formula.

\begin {eqnarray}
   \sum {1\over i^{3/2} j^{3/2} (i+j)^{1/2}}
   &\simeq& \phantom{-}2.416~942~200 
\\
   \sum_{ij} {(i+j)^{1/2}-i^{1/2}-j^{1/2}\over i^{3/2}j^{3/2}}
   &\simeq& -8.215~157~561
\\
   \sum_{ijk} {1\over (ijk)^{1/2} (i+k)(j+k)(i+j+k)^{1/2}}
   &\simeq& \phantom{-}2.211~1
\\
   \sum_{ijk} {1\over (ij)^{3/2} k^{1/2}} \left(
              {1 \over (i+j+k)^{1/2}}
              - {1\over k^{1/2}}
           \right)
   &\simeq& -16.70
\end {eqnarray}


\section {Derivation of the sunset diagram}
\label {app:sunset}

Using the methods outlined in Appendix A, we will now reduce
the sunset diagram (\ref{eq:Pid}) to the integral representation
(\ref{eq:Pidn}) in terms of distribution functions.
Starting with
\begin {equation}
   \Pi_\psi^{\rm(d)}(0) =
      - {1\over2} \left(8\pi a\over m\right)^2
      \sumint_{QKL}
        { \beta \, \delta_{l_0-q_0-k_0} \, (2\pi)^d \delta^{(d)}(\l-\q-\k)
         \over
         (i l_0 + \omega_l) (i q_0 + \omega_q) (i k_0 + \omega_k) } \,,
\label {eq:flop}
\end {equation}
and rewriting the Kronecker $\delta$ function in (\ref{eq:flop})
as in Appendix A, we obtain
\begin {equation}
   \Pi_\psi^{\rm(d)}(0) =
   - {1\over2} \left(8\pi a\over m\right)^2
     \int_0^\beta d\alpha
       \sumint_L {e^{-i\alpha l_0}\over il_0 + \omega_p}
       \sumint_Q {e^{+i\alpha q_0}\over iq_0 + \omega_q}
       \sumint_K {e^{+i\alpha k_0}\over ik_0 + \omega_k}
       \, (2\pi)^d \delta^{(d)}(\l-\q-\k)
   \,.
\end {equation}
Using the frequency sums (\ref{eq:fsums}) and then performing the
$\alpha$ integration,
\begin {equation}
   \Pi_\psi^{\rm(d)}(0) =
   - {1\over2} \left(8\pi a\over m\right)^2
     \int_{\q\k\l} n(\omega_l)\, n(\omega_q)\, n(\omega_k)\,
       {e^{\beta\omega_l} - e^{\beta(\omega_q+\omega_k)} \over
       \omega_l - \omega_q - \omega_k }
       \, (2\pi)^d \delta^{(d)}(\l-\q-\k)
   \,.
\end {equation}
The integrand is well behaved at $\omega_l = \omega_q+\omega_k$
(except for the infrared divergence where $l$, $q$, and $k$ all go
to zero, which is dealt with in the main text).  However, as in
Appendix A, it is convenient to introduce a spurious principal part
prescription at this stage.  Then, using (\ref{eq:nexp}) and
permuting integration variables, one arrives at (\ref{eq:Pidn}).

The last term in (\ref{eq:Pidn}), involving just one $n$, is proportional to
\begin {equation}
      \int_{\q\k\l} \PP
         {n(\omega_l)
         \over \omega_l - \omega_q - \omega_k}
         \, (2\pi)^d \delta^{(d)}(\l-\q-\k) .
\end {equation}
The $\q\k$ part of this integration is just a special case of
(\ref{eq:zero}) with the momentum labels changed and $\p$ set to zero.
As described in Appendix A, it therefore gives zero contribution in
dimensional regularization for $d=3$.


\end {document}